\pdfoutput=1

\documentclass[11pt,twoside,a4paper,cmspaper,final,collab]{cms-tdr}
\def\svnVersion{7dc3138-D}\def\svnDate{2019/11/05}\def\cmsCernNoTag{CERN-EP-2019-171}\def\cmsCernDate{\today}\def\cmsMessage{"Published in the Journal of High Energy Physics as \href{http://dx.doi.org/10.1007/JHEP11(2019)109}{\doi{10.1007/JHEP11(2019)109}.}"}

\begin{document}\cmsNoteHeader{SUS-18-007}

\hyphenation{had-ron-i-za-tion}
\hyphenation{cal-or-i-me-ter}
\hyphenation{de-vices}
\RCS$HeadURL$
\RCS$Id$
\newlength\cmsFigWidth
\ifthenelse{\boolean{cms@external}}{\setlength\cmsFigWidth{0.49\textwidth}}{\setlength\cmsFigWidth{0.65\textwidth}}
\ifthenelse{\boolean{cms@external}}{\providecommand{\cmsLeft}{upper\xspace}}{\providecommand{\cmsLeft}{left\xspace}}
\ifthenelse{\boolean{cms@external}}{\providecommand{\cmsRight}{lower\xspace}}{\providecommand{\cmsRight}{right\xspace}}

\newlength\cmsTabSkip\setlength{\cmsTabSkip}{1ex}

\newcommand{\hbb}{\ensuremath{\PH\to \cPqb\cPaqb}\xspace}
\newcommand{\hgg}{\ensuremath{\PH\to \gamma\gamma}\xspace}
\newcommand{\mgg}{\ensuremath{m_{\gamma\gamma}}\xspace}
\newcommand{\MR}{\ensuremath{M_{\mathrm{R} }\xspace}}
\newcommand{\Rtwo}{\ensuremath{R^2}\xspace}
\newcommand{\MRT}{\ensuremath{M^{\mathrm{R}}_{\mathrm{T} }\xspace}}
\providecommand{\PSGcmpDo}{\ensuremath{\widetilde{\chi}^\mp_{1}}\xspace}
\newcommand{\ptOverM}{\ensuremath{\pt^{\gamma\gamma} / m_{\gamma\gamma}}\xspace}
\providecommand{\cmsTable}[1]{\resizebox{\textwidth}{!}{#1}}

\cmsNoteHeader{SUS-18-007}
\title{Search for supersymmetry using Higgs boson to diphoton decays at $\sqrt{s} = 13$\TeV}

\date{\today}

\abstract{
        A search for supersymmetry (SUSY) is presented where at least one Higgs boson is produced and
decays to two photons in the decay chains of pair-produced SUSY particles. Two analysis
strategies are pursued: one focused on strong SUSY production and the other focused on electroweak
SUSY production. The presence of charged leptons, additional Higgs boson candidates, and various
kinematic variables are used to categorize events into search regions that are sensitive to
different SUSY scenarios. The results are based on data from proton-proton collisions at the Large
Hadron Collider at a center-of-mass energy of 13\TeV collected by the CMS experiment, corresponding
to an integrated luminosity of 77.5\fbinv. No statistically significant excess of events is observed
relative to the standard model expectations. We exclude bottom squark pair production for bottom
squark masses below 530\GeV and a lightest neutralino mass of 1\GeV; wino-like chargino-neutralino
production in gauge-mediated SUSY breaking (GMSB) for chargino and neutralino masses 
below 235\GeV with a gravitino mass of 1\GeV; and higgsino-like chargino-neutralino production in
GMSB, where the neutralino decays exclusively to a Higgs 
boson and a gravitino for neutralino masses below 290\GeV.
}

\hypersetup{
pdfauthor={CMS Collaboration},
pdftitle={Search for supersymmetry using Higgs boson to diphoton decays at sqrt{s} = 13 TeV},
pdfsubject={CMS},
pdfkeywords={CMS, physics, SUSY, Higgs}}

\maketitle

\section{Introduction}
\label{sec:intro}

The Higgs boson ($\PH$) provides an intriguing opportunity to explore physics beyond the 
standard model (SM) of particle physics. Many scenarios of physics beyond the SM postulate 
the existence of cascade decays of heavy states involving Higgs 
bosons~\cite{Monaco:2013poa,Duarte:2017bbq}. In minimal supersymmetry (SUSY)~\cite{Dimopoulos:1981zb}, 
a Higgs boson may appear in processes involving the bottom squark ($\PSQb$), the SUSY 
partner of the bottom quark. Bottom squarks are produced via strong interactions and then 
may decay to a Higgs boson, quarks, and the lightest SUSY particle (LSP). Similarly charginos 
or neutralinos produced through the electroweak interaction may decay to a Higgs boson and 
the LSP. Of particular interest are gauge-mediated SUSY breaking (GMSB) scenarios, where the 
lightest neutralino may decay to a Higgs boson and the gravitino 
LSP~($\PXXSG$)~\cite{Dimopoulos:1996vz,Matchev:1999ft}. Similar searches have been performed by the 
ATLAS and CMS Collaborations using proton-proton (pp) collisions
at the CERN LHC at center-of-mass energies of 8~\cite{Aad:2015jqa,Khachatryan:2014mma}
and 13\TeV~\cite{Sirunyan:2017eie,Aaboud:2018ngk,ATLAS:2019mge,Aaboud:2018htj}.

We search for evidence of SUSY that produces an excess of events with 
one or more Higgs bosons decaying to two photons and large missing transverse
momentum using pp collision data collected by the CMS experiment at the LHC at a center-of-mass
energy of 13\TeV in 2016 and 2017, corresponding to an integrated luminosity of 77.5\fbinv.
Kinematic variables that discriminate the SUSY signal from SM backgrounds are used to separate 
events into several mutually exclusive categories, and the diphoton mass from 
the $\PH\to\gamma\gamma$ decay is used to extract the signal
from the background. The branching ratio for $\PH\to\gamma\gamma$ of $0.227\%$ 
from the SM is assumed. The dominant backgrounds are SM production of diphoton and
photon+jets, which are modeled by functional fits to the diphoton mass distribution.
The SM Higgs boson background constitutes a small fraction of the background
for most of the phase space used in the search and is estimated from
simulation samples.

\begin{figure*}[h]
\centering
\includegraphics[width=0.4\textwidth]{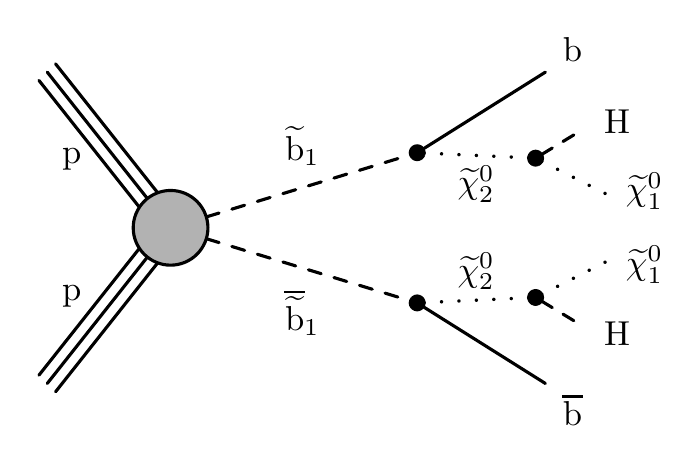}
\includegraphics[width=0.4\textwidth]{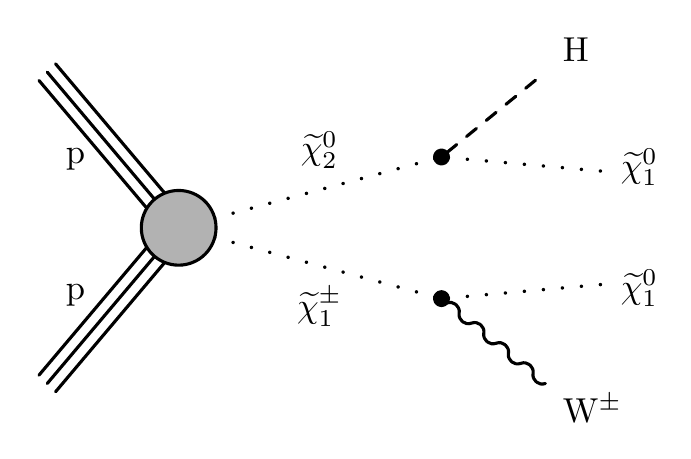}\\
\includegraphics[width=0.4\textwidth]{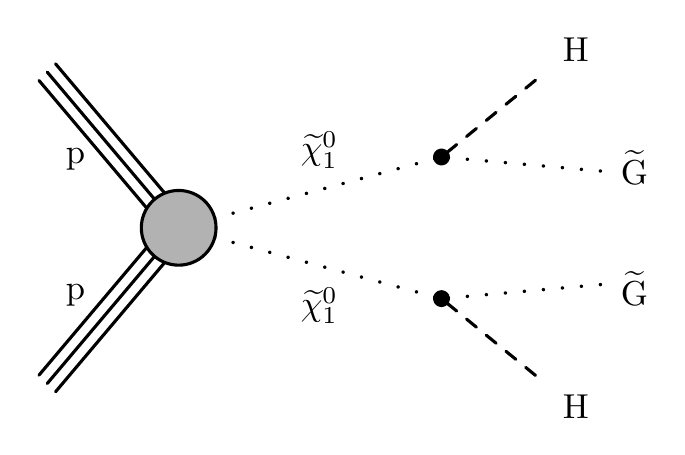}
\includegraphics[width=0.4\textwidth]{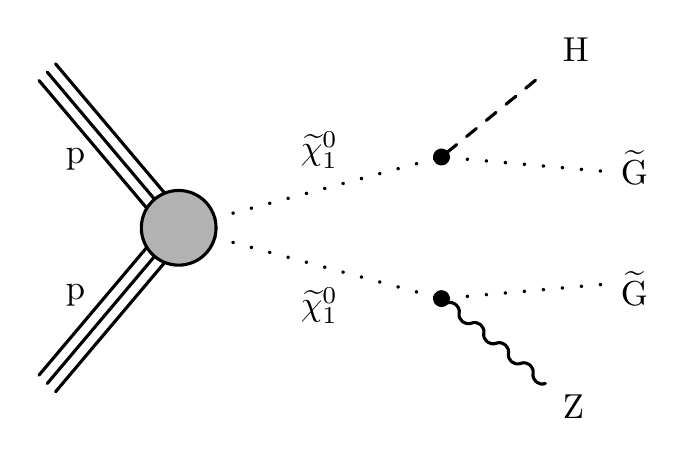}
\caption{Diagrams displaying the simplified models that are being considered. Upper left: bottom squark pair production; upper right: wino-like chargino-neutralino production; lower: the two relevant decay modes for higgsino-like neutralino pair production in the GMSB scenario.
\label{fig:SMSDiagrams}}
\end{figure*}

We have designed a new analysis to extend our sensitivity to both strong and electroweak SUSY production
over the previously published result~\cite{Sirunyan:2017eie}.
Two analysis strategies are pursued: one focuses on the electroweak production of charginos
and neutralinos by introducing additional event categories containing one or two charged-lepton candidates,
thereby enhancing the sensitivity to SUSY signatures involving \PW and \PZ bosons, and the other is optimized
for strong production by categorizing events in the number of jets and the number of jets identified
as originating from the fragmentation of \cPqb quarks (``\cPqb-tagged").
The use of the two strategies enhances the overall sensitivity of the search, and increases
the robustness of the result by exploring alternative phase space regions. Finally, we interpret the
results in various simplified model scenarios of SUSY as summarized in Fig.~\ref{fig:SMSDiagrams},
including bottom squark pair production, chargino-neutralino, and neutralino-pair production.

In this paper, we discuss the CMS detector in Section~\ref{sec:detector}, the event
simulation in Section~\ref{sec:simulation}, the event reconstruction and selection
in Section~\ref{sec:evtsel}, the analysis strategy in Section~\ref{sec:strategy},
the background estimation in Section~\ref{sec:bkg}, the systematic uncertainties
in Section~\ref{sec:systematics}, and the results and interpretations in Section~\ref{sec:results}.
A summary is given in Section~\ref{sec:summary}.

\section{The CMS detector}
\label{sec:detector}
The central feature of the CMS detector is a superconducting solenoid of 6\unit{m} internal diameter, providing a magnetic field of 3.8\unit{T}. Within the solenoid volume are a silicon pixel and strip tracker, a lead tungstate crystal electromagnetic calorimeter (ECAL), and a brass and scintillator hadron calorimeter, each composed of a barrel and two endcap sections. Forward calorimeters extend the pseudorapidity ($\eta$) coverage provided by the barrel and endcap detectors. Muons are measured in gas-ionization detectors embedded in the steel flux-return yoke outside the solenoid. The first level of the CMS trigger system~\cite{Khachatryan:2016bia}, composed of custom hardware processors, uses information from the
calorimeters and muon detectors to select the most interesting events in a fixed time interval of less than 4\mus. The high-level
trigger processor farm further decreases the event rate from around 100\unit{kHz} to less than 1\unit{kHz} before data storage.
A more detailed description of the CMS detector, together with a definition of the coordinate system used and the relevant
kinematic variables, can be found in Ref.~\cite{Chatrchyan:2008zzk}.

\section{Event simulation}\label{sec:simulation}

Simulated Monte Carlo (MC) event samples are used to model the SM Higgs boson backgrounds and the SUSY signal models.
Simulated samples of SM Higgs boson production through gluon fusion, vector boson fusion, associated production
with a $\PW$ or a $\cPZ$ boson, $\bbbar\PH$, and $\ttbar\PH$ are generated using the next-to-leading
order (NLO) \MGvATNLO v2.2.2~\cite{Alwall:2014hca} event generator. The Higgs boson mass is assumed to be 125\GeV
for the simulated event samples and is within the uncertainty of the currently best measured
value~\cite{Aad:2015zhl,Sirunyan:2017exp}. The Higgs boson production cross sections are taken from
Ref.~\cite{deFlorian:2016spz} and are computed to next-to-next-to-leading order plus next-to-next-to-leading
logarithm in the quantum chromodynamics (QCD) coupling constant and to NLO in the electroweak coupling constant.
For the gluon fusion production mode, the sample is generated with up to two extra partons from
initial-state radiation (ISR) at NLO accuracy and uses the FxFx matching scheme described
in Ref.~\cite{Frederix:2012ps}. The SUSY signal MC samples are generated
using \MGvATNLO at leading order accuracy with up to two extra partons in the matrix element calculations, with
the MLM matching scheme described in Ref.~\cite{Alwall:2007fs}. For samples simulating the 2016
data set, \PYTHIA v8.212~\cite{Sjostrand:2014zea} is used to model the fragmentation and
parton showering with the CUETP8M1 tune~\cite{Skands2014}, while for samples simulating the 2017
data set, \PYTHIA v8.226 is used with the CP5~\cite{CMS-PAS-GEN-17-001} tune.
The {NNPDF3.0}~\cite{Ball:2014uwa}~and~{NNPDF3.1}~\cite{Ball:2017nwa} parton distribution
function (PDF) sets are used for the 2016 and 2017 simulation samples, respectively.
The production cross section for squark pair production is computed at NLO plus
next-to-leading logarithmic (NLL) accuracy in QCD~\cite{NLONLL1,NLONLL2,NLONLL3,NLONLL4,NLONLL5,Borschensky:2014cia}
under the assumption that all SUSY particles other than those in the relevant diagram are too heavy to participate
in the interaction. The cross sections for higgsino pair production are computed
at NLO+NLL precision in the limit of mass-degenerate higgsinos $\PSGczDt$, $\PSGcpmDo$,
and $\PSGczDo$, with all the other sparticles assumed to be heavy and
decoupled~\cite{Beenakker:1999xh,Fuks:2012qx,Fuks:2013vua}. Following the convention
of real mixing matrices and signed neutralino or chargino masses~\cite{Skands:2003cj},
we set the mass of $\PSGczDo$ ($\PSGczDt$) to positive (negative) values. The product
of the third and fourth elements of the corresponding rows of the neutralino mixing
matrix $N$ is $+0.5$ ($-0.5$). The elements $U_{12}$ and $V_{12}$ of the chargino
mixing matrices are set to 1.

The SM Higgs boson background samples are simulated using a \GEANTfour-based model~\cite{geant4} of the CMS 
detector. To cover the large SUSY signal parameter space in reasonable computation time, the signal model 
samples are simulated with the CMS fast simulation package~\cite{fastsim,Giammanco:2014bza}, which has been 
validated to produce accurate predictions of object identification efficiencies and momentum resolution. 
All simulated events include the effects of additional pp interactions in the same or adjacent beam bunch 
crossings (pileup), and are processed with the same chain of reconstruction programs used for collision data.

To improve the \MADGRAPH modeling of ISR in the SUSY signal MC samples, we apply a shape correction as a function of the
multiplicity of ISR jets for bottom squark pair production and as a function of the transverse momentum ($\pt^{\mathrm{ISR}}$)
of the chargino-neutralino system for chargino-neutralino production, derived from studies of \ttbar and \PZ+jets
events, respectively~\cite{Chatrchyan:2013xna}. The correction factors vary 
between 0.92 and 0.51 for the ISR jet multiplicity between one
and six, and between 1.18 and 0.78 for $\pt^{\mathrm{ISR}}$ between 125 and 600\GeV. The corrections have a
small effect on the signal yields for all the simplified models considered at the level of about $1\%$. 
For the bottom squark pair production signal model,
the full effect of the correction is propagated as a systematic uncertainty.
For the chargino-neutralino production one half of effect of the correction
is propagated as a systematic uncertainty.

\section{Event reconstruction and selection}
\label{sec:evtsel}

The search with the 2016 data set uses events selected by the diphoton high-level trigger, which requires two photons with \pt
above 30~and~18\GeV for the leading and subleading photons, respectively. 
For the 2017 data set, to cope with the increased instantaneous
luminosity, the \pt requirement on the subleading photon was increased to 22\GeV in order to reduce the trigger rate.
The efficiency of the trigger for events with two identified photons is above $98\%$.

Events are reconstructed using the CMS particle flow (PF) algorithm~\cite{Sirunyan:2017ulk},
which uses the information from the tracker, calorimeter, and muon systems to construct an
optimized global description of the event. The reconstructed vertex with the largest value of summed physics-object $\pt^2$ is 
taken to be the primary interaction vertex. The physics objects used in this context 
are the objects returned by a jet finding algorithm~\cite{antikt,Cacciari:2011ma} 
applied to all charged tracks associated with the vertex under 
consideration, plus the corresponding associated missing transverse momentum.

As the signal is predominantly produced in the central region
of the detector, we select events with at least two photons reconstructed in the barrel
region ($\abs{\eta}<1.44$). The measured energy of photons is corrected for clustering 
and local geometric effects using an energy regression trained on Monte Carlo (MC) 
simulation, and calibrated using a combination of $\Pgpz\to\Pgg\Pgg$, $\Pgh\to\Pgg\Pgg$, and $\PZ\to\Pe\Pe$
candidates~\cite{Khachatryan:2015iwa}. The regression also provides an estimate of the uncertainty of the 
energy measurement that is used to separate events into high- and low-resolution categories.
The photons are required to satisfy the photon identification requirements based on
electromagnetic shower shape, hadronic to electromagnetic energy ratio, and
isolation around the photon candidate. A photon is considered isolated if the \pt sum of the PF candidates
from charged and neutral hadrons and photons within a cone of $0.3$ in
$\Delta \mathrm{R}=\sqrt{\smash[b]{(\Delta\eta)^2+ (\Delta\phi)^2}}$,
where $\phi$ is the azimuthal angle in radians,
are each below a set threshold. The isolation sums are corrected for the effect of pileup by
subtracting the average energy deposited as estimated by the pileup energy density $\rho$~\cite{Cacciari:2007fd}.
If the photon is matched to a reconstructed electron that is inconsistent with a conversion candidate,
it is discarded. A loose working point is used for the photon identification, which has
an efficiency of approximately 90\%, uniform in \pt and $\eta$. The leading (subleading) photon
is required to have $\pt / \mgg > 0.33$ (0.25), where \mgg is the reconstructed diphoton mass.
The diphoton mass is required to be larger than $100$\GeV. The two photons with the largest \pt,
selected according to the identification criteria above, are considered to be the decay products
of the Higgs boson candidate.

The PF candidates are clustered into jets using the anti-\kt algorithm~\cite{antikt,Cacciari:2011ma} with a
distance parameter of $0.4$.
Jet energy corrections are applied and derived based
on a combination of simulation studies, accounting for the nonlinear detector response and
the presence of pileup, together with in-situ measurements of the energy balance in dijet and
$\gamma$+jet events using the methods described in Ref.~\cite{Khachatryan:2016kdb}.
Jets originating from a heavy-flavor parton are identified by the
combined secondary vertex (CSVv2) tagger algorithm~\cite{btag} using a loose
working point. The resulting efficiency is about $80\%$, while the mistag rate for light-quark and gluon jets is approximately 10\%.
We identify each jet with $\pt > 20$\GeV that satisfies the loose working point as a \cPqb-tagged jet.
Other jets with $\pt>30$\GeV and $\abs{\eta}<2.4$ are considered in this analysis for the purpose of
jet counting. Electrons and muons in the region $\abs{\eta}<2.4$
and with $\pt>20$\GeV are selected from the PF candidates, and a loose identification
working point is used. Jets that overlap with the selected electrons, muons, and photons in a cone of size $\Delta\mathrm{R}=0.4$
are discarded. Electrons in a cone of size $\Delta\mathrm{R}=1.0$ and muons in a cone of size $\Delta\mathrm{R}=0.5$
around the selected photons are discarded. A larger veto cone is used for electrons to suppress photon conversions.

The transverse component of the negative vectorial sum of the momenta of all PF candidates is the missing transverse
momentum \ptvecmiss, and its magnitude is defined as \ptmiss.
Dedicated filters~\cite{Chatrchyan:2011tn} reject events with possible beam halo
contamination or anomalous noise in the calorimeter systems that can give rise to a large \ptmiss.

\section{Analysis strategy}
\label{sec:strategy}

Two analysis strategies are pursued that employ two alternative event
categorization schemes: one focused on electroweak production (EWP analysis) of charginos and
neutralinos; and another focused on strong production (SP analysis) of bottom squarks.
For both strategies, we define event categories based on the \pt of the diphoton Higgs boson candidate, and
the presence of additional $\PZ$, $\PW$, or \hbb candidates. Within each event category, we define search region bins
based on the number of jets and \cPqb-tagged jets, and the values of kinematic variables that discriminate
between SUSY signal and SM backgrounds events. Finally, to test specific SUSY simplified model hypotheses,
we perform an unbinned extended maximum likelihood fit to the diphoton mass distribution,
simultaneously in all of the search bins defined for each analysis. 

The dominant background results from SM production of diphoton or photon+jets, and is
collectively referred to as the nonresonant background. This background exhibits a regular
falling shape as validated in the MC simulation samples, and is modeled with a fit to a family of falling
functions independently in each search region bin as described in the next section. The SM Higgs boson background
and the SUSY signal model under test exhibit a resonant shape in the diphoton mass and are constrained to the MC simulation
predictions within uncertainties. A more detailed discussion of the background fit model and the systematic
uncertainties can be found in Sections~\ref{sec:bkg}~and~\ref{sec:systematics}, respectively.

In the EWP approach, we build upon the strategy employed in a previous publication~\cite{Sirunyan:2017eie}, which categorized
events according to the \pt of the diphoton Higgs boson candidate, the presence of an additional Higgs boson candidate,
the estimated diphoton mass resolution, and the values of the ``razor'' kinematic variables~\cite{Rogan:2010kb,razor2015}.
In addition, we add event categories with one or two identified leptons, and further
optimize the binning in the kinematic variables for the enlarged data set. 
The bin boundaries have been chosen to yield the best expected signal significance
as estimated using simulation predictions of the signal and background yields.
These enhancements improve the signal
sensitivity to electroweak production of charginos and neutralinos. By isolating events with a $\PZ$, $\PW$,
or \hbb candidate in addition to the \hgg candidate, we improve the sensitivity to the simplified signal models
shown in Fig.~\ref{fig:SMSDiagrams}.

The Higgs boson candidate and any additional identified leptons or jets are clustered into two hemispheres (megajets) according to
the razor megajet algorithm~\cite{razor2015}, which minimizes the sum of the squared-invariant-mass values of the
two megajets. In order to form two hemispheres, we require that events have at least one identified lepton or jet
in addition to the Higgs boson candidate. The razor variables~\cite{Rogan:2010kb,razor2015} $\MR$ and $\Rtwo$ are
then computed as follows:
\begin{align}
 \label{eq:MRstar}
 \MR &\equiv \sqrt{(\abs{\vec{p}{}^{\,\mathrm{j}_{1}}}+\abs{\vec{p}{}^{\,\mathrm{j}_{2}}})^2 - ({p}_z{}^{\mathrm{j}_{1}}+{p}_z{}^{\mathrm{j}_{2}})^2},\\
\Rtwo &\equiv \left( \frac{\MRT}{\MR} \right)^2,\label{eq:Rtwo}
\end{align}
where $\vec{p}$ is the momentum of a megajet, $p_z$ is its longitudinal component,
and $\mathrm{j}_{1}$ and $\mathrm{j}_{2}$ are used to label the two megajets. In the definition of $\Rtwo$,
the variable $\MRT$ is defined as:
\begin{equation}
\MRT \equiv \sqrt{ \frac{\ptmiss(\pt{}^{\mathrm{j}_{1}}+\pt{}^{\mathrm{j}_{2}}) - \ptvecmiss \cdot (\ptvec{}^{\,\mathrm{j}_{1}}+\ptvec{}^{\,\mathrm{j}_{2}}) }{2}}.\label{eq:MRT}
\end{equation}
The razor variables $\MR$ and $\Rtwo$ provide discrimination between SUSY signal models and SM background processes,
with SUSY signals typically having large values of $\MR$ and $\Rtwo$, while the SM diphoton and photon+jets backgrounds
exhibit a falling spectrum in each variable.

The selected events are first categorized according to the number of electrons or muons.
Events with two same-flavor opposite-sign leptons are placed
in the ``Two-Lepton'' category if the dilepton mass satisfies the constraint $\abs{m_{\cPZ} - m_{\ell\ell}} \leq 20$\GeV. Among the
remaining events, those with at least one muon (electron) are placed in the ``Muon'' (``Electron'') category, with the
Muon category taking precedence. Events in the Electron and
Muon categories are further subdivided into the ``High-$\pt$'' and ``Low-$\pt$'' subcategories depending on whether the $\pt$ of
the Higgs boson candidate is larger or smaller than 110\GeV. For events which do not have any leptons, we search for pairs
of \cPqb-tagged jets, whose mass is between 95 and 140\GeV, and place them into the ``$\PH\PQb\cPaqb$'' category. If no such jet-pairs are
found, then we search for pairs of \cPqb-tagged jets whose mass is between 60 and 95\GeV, and place them into
the ``$\PZ\PQb\cPaqb$'' category. Events in the $\PH\PQb\cPaqb$ and $\PZ\PQb\cPaqb$ categories are also
further subdivided into the High-$\pt$ and Low-$\pt$ subcategories using the same criteria stated above.
Among the remaining events, those with the $\pt$ of the Higgs boson candidate larger than 110\GeV
are placed in the High-$\pt$ category. Finally, the
remaining events are categorized as ``High-Res'' or ``Low-Res'' if the diphoton mass resolution estimate $\sigma_{m}/m$ is
smaller or larger than 0.85\%, respectively, with $\sigma_{m}$ defined as:
\begin{equation}
\sigma_{m} = \frac{1}{2}\,\sqrt{(\sigma_{E\gamma 1}/E_{\gamma 1})^{2} + (\sigma_{{E}\gamma 2}/E_{\gamma 2})^{2}},
\end{equation}
where $E_{\gamma 1,2}$ is the energy of each photon and $\sigma_{{E}\gamma 1,2}$ is the estimated energy resolution for each photon.
The choice of the 0.85\% threshold was made to be identical to past results~\cite{Sirunyan:2017eie}, which 
was previously optimized for signal to background discrimination.

The leptonic categories select SUSY events containing decays to \PW or \PZ bosons; the $\PH\PQb\cPaqb$ ($\PZ\PQb\cPaqb$)
categories select events that contain an additional Higgs (\PZ) boson, which decays to a pair of \cPqb~jets;
the High-$\pt$ category selects SUSY events producing high-\pt Higgs bosons; and the separation into the High-Res and
Low-Res categories further improves the discrimination between any signal containing an \hgg candidate and non-resonant 
background in the remaining event sample. Finally, to distinguish SUSY signal events 
from the SM background, each event category is further divided into bins
in the $\MR$ and $\Rtwo$ variables, provided there are a sufficient number of data events in the diphoton mass sideband
to be able to estimate the background. These bins define the exclusive search regions.
For all categories except the Two-Lepton category, we impose the requirement $\MR>150$\GeV to suppress the SM backgrounds.

In the SP approach, we optimize the event categorization for strong production of bottom squark pairs, which typically
produce a larger number of jets and \cPqb-tagged jets. An alternative clustering algorithm is employed,
following Ref.~\cite{MT2at13TeV}, to produce two hemispheres referred to as pseudojets, and the kinematic
variable \mTii~\cite{MT2variable} is calculated as
\begin{equation}
  \mTii = \min_{ \ptvec^{\,\mathrm{miss X(1)}} + \ptvec^{\,\mathrm{miss X(2)}}  = \ptvecmiss}
  \left[ \max \left( \mT^{(1)} , \mT^{(2)} \right) \right],
  \label{eq.MT2.definition}
 \end{equation}
where $\ptvec^{\,\mathrm{miss X(i)}}$ (with $i$=1,2) are trial vectors obtained by
decomposing \ptvecmiss and $\mT^{(i)}$, the transverse masses obtained by pairing any of these trial
vectors with one of the two pseudojets. The minimization is performed over all trial momenta
satisfying the \ptvecmiss constraint. The \ptOverM and \mTii kinematic variables are used to enhance
the discrimination between the SUSY signal and the SM background. Two bins in the \mTii variable are
used: $\mTii<30$ and $\mTii \geq 30$\GeV;
and three bins in \ptOverM: 0--0.6, 0.6--1.0 and $\ge$1.0.

Events are also separated into the Two-Lepton, Muon, Electron, $\PH\PQb\cPaqb$, and $\PZ\PQb\cPaqb$ categories
following the same procedure as described above for the EWP approach.
The remaining events are separated into the hadronic categories depending on the number of
jets and \cPqb-tagged jets. Within each of the event categories, the exclusive search region
bins are then defined based on the values of the \ptOverM and \mTii observables.

A summary of the 35 search region bins is shown in Table~\ref{tab:BinsRazor} for the EWP analysis
and of the 64 search region bins in Tables~\ref{tab:SPBins1}~and~\ref{tab:SPBins2} for the SP analysis.

\begin{table*}[htb!]
\centering
\topcaption{A summary of the search region bins used in the EWP analysis. Events are separated into
categories based on the number of leptons, the presence of $\hbb$ candidates, the $\pt$ of the $\hgg$
candidate, and the estimated diphoton mass resolution. The High-Res and Low-Res categories are defined by
the estimated diphoton resolution mass $\sigma_{m}/m$ being smaller or larger than $0.85\%$, respectively.
For the Two-Lepton category, ``No req.'' means that no requirements are placed on the given observables.
}
\begin{tabular}{ccccc}
\hline
Bin number              & Category                & $\pt^{\gamma\gamma}$ (\GeVns{}) & $\MR$ (\GeVns{})    & $\Rtwo$              \\
\hline
EWP 0                   & Two-Lepton              & No req.                    & No req.              & No req.               \\
EWP 1                   & Muon High-$\pt$             & $\ge$110                & $\ge$150          & $\ge$0.0           \\
EWP 2                   & Muon Low-$\pt$              & 0--110                  & $\ge$150          & $\ge$0.0           \\
EWP 3                   & Electron High-$\pt$         & $\ge$110                & $\ge$150          & $\ge$0.0           \\
EWP 4                   & Electron Low-$\pt$          & 0--110                  & $\ge$150          & 0.000--0.055       \\
EWP 5                   & Electron Low-$\pt$          & 0--110                  & $\ge$150          & 0.055--0.125       \\
EWP 6                   & Electron Low-$\pt$          & 0--110                  & $\ge$150          & $\ge$0.125         \\
EWP 7                   & $\PH\PQb\cPaqb$ High-$\pt$  & $\ge$110                & $\ge$150          & 0.000--0.080       \\
EWP 8                   & $\PH\PQb\cPaqb$ High-$\pt$  & $\ge$110                & $\ge$150          & $\ge$0.080         \\
EWP 9                   & $\PH\PQb\cPaqb$ Low-$\pt$  & 0--110                   & $\ge$150          & 0.000--0.080       \\
EWP 10                  & $\PH\PQb\cPaqb$ Low-$\pt$  & 0--110                   & $\ge$150          & $\ge$0.080         \\
EWP 11                  & $\PZ\PQb\cPaqb$ High-$\pt$  & $\ge$110                & $\ge$150          & 0.000--0.035       \\
EWP 12                  & $\PZ\PQb\cPaqb$ High-$\pt$  & $\ge$110                & $\ge$150          & 0.035--0.090       \\
EWP 13                  & $\PZ\PQb\cPaqb$ High-$\pt$  & $\ge$110                & $\ge$150          & $\ge$0.090         \\
EWP 14                  & $\PZ\PQb\cPaqb$ Low-$\pt$  & 0--110                   & $\ge$150          & 0.000--0.035       \\
EWP 15                  & $\PZ\PQb\cPaqb$ Low-$\pt$  & 0--110                   & $\ge$150          & 0.035--0.090       \\
EWP 16                  & $\PZ\PQb\cPaqb$ Low-$\pt$  & 0--110                   & $\ge$150          & $\ge$0.090         \\
EWP 17                  & High-$\pt$                 & $\ge$110                 & $\ge$150          & $\ge$0.260         \\
EWP 18                  & High-$\pt$                 & $\ge$110                 & 150--250          & 0.170--0.260       \\
EWP 19                  & High-$\pt$                 & $\ge$110                 & $\ge$250          & 0.170--0.260       \\
EWP 20                  & High-$\pt$                 & $\ge$110                 & $\ge$150          & 0.000--0.110       \\
EWP 21                  & High-$\pt$                 & $\ge$110                 & 150--350          & 0.110--0.170       \\
EWP 22                  & High-$\pt$                 & $\ge$110                 & $\ge$350          & 0.110--0.170       \\
EWP 23                  & High-Res                & 0--110                   & $\ge$150          & $\ge$0.325         \\
EWP 24                  & High-Res                & 0--110                   & $\ge$150          & 0.285--0.325       \\
EWP 25                  & High-Res                & 0--110                   & $\ge$150          & 0.225--0.285       \\
EWP 26                  & High-Res                & 0--110                   & $\ge$150          & 0.000--0.185       \\
EWP 27                  & High-Res                & 0--110                   & 150--200          & 0.185--0.225       \\
EWP 28                  & High-Res                & 0--110                   & $\ge$200          & 0.185--0.225       \\
EWP 29                  & Low-Res                 & 0--110                   & $\ge$150          & $\ge$0.325         \\
EWP 30                  & Low-Res                 & 0--110                   & $\ge$150          & 0.285--0.325       \\
EWP 31                  & Low-Res                 & 0--110                   & $\ge$150          & 0.225--0.285       \\
EWP 32                  & Low-Res                 & 0--110                   & $\ge$150          & 0.000--0.185       \\
EWP 33                  & Low-Res                 & 0--110                   & 150--200          & 0.185--0.225       \\
EWP 34                  & Low-Res                 & 0--110                   & $\ge$200          & 0.185--0.225       \\
\hline
\end{tabular}
\label{tab:BinsRazor}
\end{table*}

\begin{table*}[htb!]
\centering
\topcaption{A summary of the search region bins in the leptonic and Higgs boson categories
used in the SP analysis, along with the requirements
on \ptOverM and \mTii. There are no explicit requirements on the number of
jets or \cPqb-tagged jets for these categories.
For the Two-Lepton category, ``No req.'' means that no requirements are placed on the given observables.
}
\begin{tabular}{ccccc}
\hline
Bin number           & Bin name                 & Category        & \ptOverM             & \mTii (\GeVns{})      \\
\hline
SP 0   & $\PZ_{\ell\ell}$                         & Two-Lepton     & No req.              & No req.               \\[\cmsTabSkip]
SP 1   & $1\PGm ~ \pt^{0}, ~ \mTii^{0} ~ $      & Muon           & 0.0--0.6             & 0--30              \\
SP 2   & $1\PGm ~ \pt^{0}, ~ \mTii^{30} ~ $     & Muon           & 0.0--0.6             & $\geq$30              \\
SP 3   & $1\PGm ~ \pt^{75}, ~ \mTii^{0} ~ $     & Muon           & 0.6--1.0             & 0--30              \\
SP 4   & $1\PGm ~ \pt^{75}, ~ \mTii^{30} ~ $    & Muon           & 0.6--1.0             & $\geq$30              \\
SP 5   & $1\PGm ~ \pt^{125}, ~ \mTii^{0} ~ $    & Muon            & $\ge$1.0            & 0--30              \\
SP 6   & $1\PGm ~ \pt^{125}, ~ \mTii^{30} ~ $    & Muon            & $\ge$1.0            & $\geq$30              \\[\cmsTabSkip]
SP 7   & $1\Pe ~ \pt^{0}, ~ \mTii^{0} ~ $       & Electron          & 0.0--0.6             & 0--30              \\
SP 8   & $1\Pe ~ \pt^{0}, ~ \mTii^{30} ~ $      & Electron          & 0.0--0.6             & $\geq$30              \\
SP 9   & $1\Pe ~ \pt^{75}, ~ \mTii^{0} ~ $      & Electron          & 0.6--1.0             & 0--30              \\
SP 10  & $1\Pe ~ \pt^{75}, ~ \mTii^{30} ~ $     & Electron          & 0.6--1.0             & $\geq$30              \\
SP 11  & $1\Pe ~ \pt^{125}, ~ \mTii^{0} ~ $     & Electron          & $\ge$1.0            & 0--30              \\
SP 12  & $1\Pe ~ \pt^{125}, ~ \mTii^{30} ~ $    & Electron          & $\ge$1.0            & $\geq$30              \\[\cmsTabSkip]
SP 13  & $\PZ\PQb\cPaqb ~ \pt^{0}, ~ \mTii^{0} ~ $      & $\PZ\PQb\cPaqb$            & 0.0--0.6             & 0--30             \\
SP 14  & $\PZ\PQb\cPaqb ~ \pt^{75}, ~ \mTii^{0} ~ $     & $\PZ\PQb\cPaqb$            & 0.6--1.0             & 0--30             \\
SP 15  & $\PZ\PQb\cPaqb ~ \pt^{125}, ~ \mTii^{0} ~ $    & $\PZ\PQb\cPaqb$           & $\ge$1.0             & 0--30             \\
SP 16  & $\PZ\PQb\cPaqb ~ \pt^{0}, ~ \mTii^{30} ~ $     & $\PZ\PQb\cPaqb$           & 0.0--0.6             & $\geq$30            \\
SP 17  & $\PZ\PQb\cPaqb ~ \pt^{75}, ~ \mTii^{30} ~ $    & $\PZ\PQb\cPaqb$           & 0.6--1.0              & $\geq$30            \\
SP 18  & $\PZ\PQb\cPaqb ~ \pt^{125}, ~ \mTii^{30} ~ $   & $\PZ\PQb\cPaqb$           & $\ge$1.0             & $\geq$30            \\[\cmsTabSkip]
SP 19  & $\PH\PQb\cPaqb ~ \pt^{0}, ~ \mTii^{0} ~ $     & $\PH\PQb\cPaqb$            & 0.0--0.6             & 0--30             \\
SP 20  & $\PH\PQb\cPaqb ~ \pt^{75}, ~ \mTii^{0} ~ $    & $\PH\PQb\cPaqb$            & 0.6--1.0             & 0--30             \\
SP 21  & $\PH\PQb\cPaqb ~ \pt^{125}, ~ \mTii^{0} ~ $   & $\PH\PQb\cPaqb$            & $\ge$1.0            & 0--30             \\
SP 22  & $\PH\PQb\cPaqb ~ \pt^{0}, ~ \mTii^{30} ~ $    & $\PH\PQb\cPaqb$            & 0.0--0.6             & $\geq$30            \\
SP 23  & $\PH\PQb\cPaqb ~ \pt^{75}, ~ \mTii^{30} ~ $   & $\PH\PQb\cPaqb$            & 0.6--1.0             & $\geq$30            \\
SP 24  & $\PH\PQb\cPaqb ~ \pt^{125}, ~ \mTii^{30} ~ $  & $\PH\PQb\cPaqb$            & $\ge$1.0             & $\geq$30            \\
\hline
\end{tabular}
\label{tab:SPBins1}
\end{table*}

\begin{table*}[htb!]
\centering
\topcaption{A summary of the search region bins in the leptonic and Higgs boson categories
used in the SP analysis, along with the requirements
on \ptOverM and \mTii. ``No req.'' means that no requirements are placed on the given observables.
}
\begin{tabular}{cccccc}
\hline
Bin number  & Bin name  & Jets        & \cPqb-tagged jets  & \ptOverM             & \mTii (\GeVns{})      \\
\hline
SP 25    & $0     \text{j}, ~ \ge$0$\cPqb,  ~ \pt^{0} ~ $     & 0          &  No req.          & 0.0--0.6             & No req.                \\
SP 26    & $0     \text{j}, ~ \ge$0$\cPqb,  ~ \pt^{75} ~ $    & 0          &  No req.          & 0.6--1.0             & No req.               \\
SP 27    & $0     \text{j}, ~ \ge$0$\cPqb,  ~ \pt^{125} ~ $   & 0          &  No req.          & $\ge$1.0            & No req.                 \\[\cmsTabSkip]
SP 28    & $1$--$3 \text{j}, ~ 0\cPqb, ~ \pt^{0}, ~ \mTii^{0} ~ $         & 1--3       &  0                & 0.0--0.6             & 0--30                   \\
SP 29    & $1$--$3 \text{j}, ~ 0\cPqb, ~ \pt^{0}, ~ \mTii^{30} ~ $        &  1--3      &  0                & 0.0--0.6             & $\geq$30                 \\
SP 30    & $1$--$3 \text{j}, ~ 0\cPqb, ~ \pt^{75}, ~ \mTii^{0} ~ $        &  1--3       &  0                & 0.6--1.0              & 0--30                   \\
SP 31    & $1$--$3 \text{j}, ~ 0\cPqb, ~ \pt^{75}, ~ \mTii^{30} ~ $       &  1--3      &  0                & 0.6--1.0              & $\geq$30                 \\
SP 32    & $1$--$3 \text{j}, ~ 0\cPqb, ~ \pt^{125}, ~ \mTii^{0} ~ $       &  1--3       &  0                & $\ge$1.0              & 0--30                  \\
SP 33    & $1$--$3 \text{j}, ~ 0\cPqb, ~ \pt^{125}, ~ \mTii^{30} ~ $      &  1--3       &  0                & $\ge$1.0              & $\geq$30                 \\
SP 34    & $1$--$3 \text{j}, ~ 1\cPqb, ~ \pt^{0}, ~ \mTii^{0} ~ $         &   1--3       &  1               & 0.0--0.6              & 0--30                    \\
SP 35    & $1$--$3 \text{j}, ~ 1\cPqb, ~ \pt^{0}, ~ \mTii^{30} ~ $        &   1--3       &  1               & 0.0--0.6              & $\geq$30                 \\
SP 36    & $1$--$3 \text{j}, ~ 1\cPqb, ~ \pt^{75}, ~ \mTii^{0} ~ $        &   1--3       &  1                & 0.6--1.0               & 0--30                 \\
SP 37    & $1$--$3 \text{j}, ~ 1\cPqb, ~ \pt^{75}, ~ \mTii^{30} ~ $       &  1--3       &  1                & 0.6--1.0               & $\geq$30                \\
SP 38    & $1$--$3 \text{j}, ~ 1\cPqb, ~ \pt^{125}, ~ \mTii^{0} ~ $       &  1--3       &  1                & $\ge$1.0               & 0--30                  \\
SP 39    & $1$--$3 \text{j}, ~ 1\cPqb, ~ \pt^{125}, ~ \mTii^{30} ~ $      &  1--3       &  1                & $\ge$1.0                 & $\geq$30               \\
SP 40    & $1$--$3 \text{j}, ~ \ge$2$\cPqb, ~ \pt^{0}, ~ \mTii^{0} ~ $     &   1--3      &  $\ge$2          & 0.0--0.6              & 0--30                   \\
SP 41    & $1$--$3 \text{j}, ~ \ge$2$\cPqb, ~ \pt^{0}, ~ \mTii^{30} ~ $    &  1--3      &  $\ge$2          & 0.0--0.6              & $\geq$30                 \\
SP 42    & $1$--$3 \text{j}, ~ \ge$2$\cPqb, ~ \pt^{75}, ~ \mTii^{0} ~ $    &  1--3      &  $\ge$2          & 0.6--1.0               & 0--30                  \\
SP 43    & $1$--$3 \text{j}, ~ \ge$2$\cPqb, ~ \pt^{75}, ~ \mTii^{30} ~ $   &  1--3      &  $\ge$2          & 0.6--1.0               & $\geq$30                 \\
SP 44    & $1$--$3 \text{j}, ~ \ge$2$\cPqb, ~ \pt^{125}, ~ \mTii^{0} ~ $   &  1--3       &  $\ge$2         & $\ge$1.0                & 0--30                  \\
SP 45    & $1$--$3 \text{j}, ~ \ge$2$\cPqb, ~ \pt^{125}, ~ \mTii^{30} ~ $  &  1--3       &  $\ge$2          & $\ge$1.0                & $\geq$30                \\ [\cmsTabSkip]
SP 46    & $\ge$4$\text{j}, ~ 0\cPqb, ~ \pt^{0}, ~ \mTii^{0} ~ $         &  $\geq$4   & 0                 & 0.0--0.6              & 0--30                     \\
SP 47    & $\ge$4$\text{j}, ~ 0\cPqb, ~ \pt^{0}, ~ \mTii^{30} ~ $        &  $\geq$4   & 0                 & 0.0--0.6              & $\geq$30                 \\
SP 48    & $\ge$4$\text{j}, ~ 0\cPqb, ~ \pt^{75}, ~ \mTii^{0} ~ $        &  $\geq$4   & 0                 & 0.6--1.0               & 0--30                   \\
SP 49    & $\ge$4$\text{j}, ~ 0\cPqb, ~ \pt^{75}, ~ \mTii^{30} ~ $       &  $\geq$4   & 0                 & 0.6--1.0               & $\geq$30                 \\
SP 50    & $\ge$4$\text{j}, ~ 0\cPqb, ~ \pt^{125}, ~ \mTii^{0} ~ $       &  $\geq$4   & 0                 & $\ge$1.0              & 0--30                   \\
SP 51    & $\ge$4$\text{j}, ~ 0\cPqb, ~ \pt^{125}, ~ \mTii^{30} ~ $      &  $\geq$4   & 0                 & $\ge$1.0                & $\geq$30                \\
SP 52    & $\ge$4$\text{j}, ~ 1\cPqb, ~ \pt^{0}, ~ \mTii^{0} ~ $         &  $\geq$4   & 1                 & 0.0--0.6              & 0--30                 \\
SP 53    & $\ge$4$\text{j}, ~ 1\cPqb, ~ \pt^{0}, ~ \mTii^{30} ~ $        &  $\geq$4   & 1                 & 0.0--0.6              & $\geq$30                \\
SP 54    & $\ge$4$\text{j}, ~ 1\cPqb, ~ \pt^{75}, ~ \mTii^{0} ~ $        &  $\geq$4   & 1                 & 0.6--1.0               & 0--30                 \\
SP 55    & $\ge$4$\text{j}, ~ 1\cPqb, ~ \pt^{75}, ~ \mTii^{30} ~ $       &  $\geq$4   & 1                 & 0.6--1.0               & $\geq$30                \\
SP 56    & $\ge$4$\text{j}, ~ 1\cPqb, ~ \pt^{125}, ~ \mTii^{0} ~ $       &  $\geq$4   & 1                 & $\ge$1.0               & 0--30                 \\
SP 57    & $\ge$4$\text{j}, ~ 1\cPqb, ~ \pt^{125}, ~ \mTii^{30} ~ $      &  $\geq$4   & 1                 & $\ge$1.0                & $\geq$30                \\
SP 58    & $\ge$4$\text{j}, ~ \ge$2$\cPqb, ~ \pt^{0}, ~ \mTii^{0} ~ $     &  $\geq$4  &  $\ge$2          & 0.0--0.6               & 0--30                    \\
SP 59    & $\ge$4$\text{j}, ~ \ge$2$\cPqb, ~ \pt^{0}, ~ \mTii^{30} ~ $    &  $\geq$4  &  $\ge$2          & 0.0--0.6               & $\geq$30                 \\
SP 60    & $\ge$4$\text{j}, ~ \ge$2$\cPqb, ~ \pt^{75}, ~ \mTii^{0} ~ $    &  $\geq$4  &  $\ge$2          & 0.6--1.0               & 0--30                    \\
SP 61    & $\ge$4$\text{j}, ~ \ge$2$\cPqb, ~ \pt^{75}, ~ \mTii^{30} ~ $   &  $\geq$4  &  $\ge$2          & 0.6--1.0               & $\geq$30                 \\
SP 62    & $\ge$4$\text{j}, ~ \ge$2$\cPqb, ~ \pt^{125}, ~ \mTii^{0} ~ $   &  $\geq$4  &  $\ge$2          & $\ge$1.0                & 0--30                   \\
SP 63    & $\ge$4$\text{j}, ~ \ge$2$\cPqb, ~ \pt^{125}, ~ \mTii^{30} ~ $  &  $\geq$4  &  $\ge$2          & $\ge$1.0                 & $\geq$30                 \\
\hline
& & & & &
\end{tabular}
\label{tab:SPBins2}
\end{table*}

Finally, to test specific SUSY simplified model hypotheses, we perform a
combined simultaneous fit using all the search regions defined for each analysis.
The final result for each signal model is obtained from the analysis with the
best expected sensitivity. The diphoton mass distribution is fit independently in each search region, while the
expected yields for the SM Higgs background and SUSY signal model among the different
search regions are constrained to the predicted values.

Search region bins with large values of $\pt^{\gamma\gamma}$ and large values of 
the kinematic variables \MR~and \mTii yield the best sensitivity
for SUSY signals with larger squark or neutralino masses, as backgrounds are heavily suppressed.
The event categories with one lepton, two leptons, a $\PZ\to\cPqb\cPaqb$ candidate, or a \hbb candidate
yield increasingly better sensitivity for more compressed regions as the neutralino mass approaches the Higgs boson
mass.

\section{Backgrounds}\label{sec:bkg}

Two types of backgrounds can be identified for this search: a nonresonant one stemming from the
SM production of diphotons or a photon and a jet, and a resonant background from SM Higgs boson production.
To model the nonresonant background, a set of possible functions
is chosen from sums of exponential functions, sums of Bernstein polynomials, Laurent series, and
sums of power-law functions. To determine the best functional form, two alternative strategies are followed for
the EWP and SP analyses. As we do not know a priori the exact shape of the background, it is important that the functional
form used is capable of adequately describing a sufficiently large range of background shapes to cover potential systematic effects that
affect the shapes. At the same time we do not want to arbitrarily increase the number of fit parameters
without yielding additional robustness against systematic uncertainties.

The EWP analysis uses the Akaike information criterion (AIC)~\cite{AIC}
to determine which functional forms are most appropriate to describe the background spectrum.
The same procedure was employed in the previous version of this
search~\cite{Sirunyan:2017eie}. Bias tests are performed by drawing random events using one functional
form and fitting the resulting pseudo-data set to another functional form. The functional form with the
best AIC measure passing the bias test is chosen to describe the nonresonant background.

For the SP analysis, the background fit is performed by discrete profiling using the "envelope" method~\cite{DiscProf}.
The background functional form is treated as a discrete nuisance parameter in the likelihood fit.
A penalty is assigned to the likelihood for each parameter in the function.
The envelope with the best likelihood is determined by the discrete profiling method taking penalties into account.
These two alternative background modeling methods were studied in a past CMS measurement of the SM Higgs process
in the diphoton decay channel and similar accuracy is expected~\cite{Khachatryan:2014ira}.

The shape of the SM Higgs boson background and the SUSY signals is modeled by a 
double Crystal Ball function~\cite{Oreglia:1980cs,Gaiser:1982yw},
fitted to the diphoton mass distribution from the MC simulation separately in each search region bin.
The parameters of each double Crystal Ball function are held constant in the signal extraction fit procedure. 
The normalization of the SM Higgs boson background in each bin is constrained to the MC simulation prediction 
to within systematic uncertainties.

\section{Systematic uncertainties}
\label{sec:systematics}

The dominant systematic uncertainties in this search are the normalization and shape of the nonresonant background
associated with the fitted functional form. They are propagated by profiling the associated unconstrained 
functional form parameters. The fraction of the total uncertainty due to the nonresonant background fit ranges from
$75\%$ to $99\%$, and is above $90\%$ for most search region bins. The subdominant systematic uncertainties in the
SM Higgs boson background and SUSY signal are propagated through independent
log-normal nuisance parameters that take both theoretical and instrumental effects into account.
These systematic uncertainties affect the event yield predictions of the SM Higgs boson background and SUSY signal
in the different search region bins, and are propagated as shape uncertainties.
The independent systematic effects considered include missing higher-order QCD corrections, PDFs, trigger
and object selection efficiencies, jet energy
scale uncertainties, \cPqb-tagging efficiency, lepton identification efficiencies, fast simulation \ptmiss
modeling, and the uncertainty in the integrated luminosity.
The typical size of these effects on the signal and background yields are summarized in Table~\ref{tab:SignalSystematics},
and are approximately the same for the SP and EWP analyses.
Systematic uncertainties due to missing higher-order corrections are estimated by the use of the
procedure outlined in Ref.~\cite{Kalogeropoulos:2018cke}, where the
factorization ($\mu_{\mathrm{F}}$) and renormalization ($\mu_{\mathrm{R}}$) scales are varied
independently by factors of 0.5 and 2.0. The PDF systematic uncertainties
are propagated for the SM Higgs background
as a shape uncertainty using the LHC4PDF procedure~\cite{Butterworth:2015oua}.

Because of the imperfect simulation of the effects of pileup and transparency loss from radiation
damage in the ECAL crystals, we observe some simulation mismodeling
of the estimated mass resolution, which can migrate events between the High-Res and Low-Res
event categories of the EWP analysis. As a result, a systematic uncertainty of $10$--$24\%$, measured
using a $\PZ\to\Pep\Pem$ control sample, is propagated to the prediction of the SM Higgs
boson background and SUSY signal yields in the High-Res and Low-Res event categories.
The systematic uncertainty in the photon energy scale
is implemented as a Gaussian-distributed nuisance parameter that shifts the Higgs boson mass
peak position, constrained in the fit to lie within approximately $1\%$ of the nominal Higgs
boson mass observed in simulation. The systematic uncertainty for the modeling of the ISR
for the signal process is also propagated.

\begin{table*}[htb]
\centering
\topcaption{Summary of systematic uncertainties on the SM Higgs boson background and signal yield predictions,
and the size of their effect on the signal yield.}
\label{tab:SignalSystematics}
\begin{tabular}{lc}
\hline
Uncertainty source & Uncertainty size (\%) \\
\hline
\multirow{3}{*}{PDFs and QCD scale variations} & $10$--$30$ (SM Higgs boson) \\
                                               & $5$--$10$ (EWK SUSY signal) \\
                                               & $15$--$30$ (Strong SUSY signal) \\
Signal ISR modeling                        & $5$--$25$ \\
$\sigma_{m}/m$ categorization               & $10$--$24$ \\
Fast simulation \ptmiss modeling           & $3$--$16$ \\
Luminosity                                 & $2.3$--$2.5$ \\
Trigger and selection efficiency           & $3$ \\
Lepton efficiency                          & $4$ \\
Jet energy scale                           & $1$--$5$ \\
Photon energy scale                        & $1$ \\
\cPqb-tagging efficiency                   & $4$ \\
$\PH\to\gamma\gamma$ branching fraction    & $2$ \\
\hline
\end{tabular}
\end{table*}

\section{Results and interpretation}
\label{sec:results}

\begin{figure}[htb]
\centering
\includegraphics[width=0.49\textwidth]{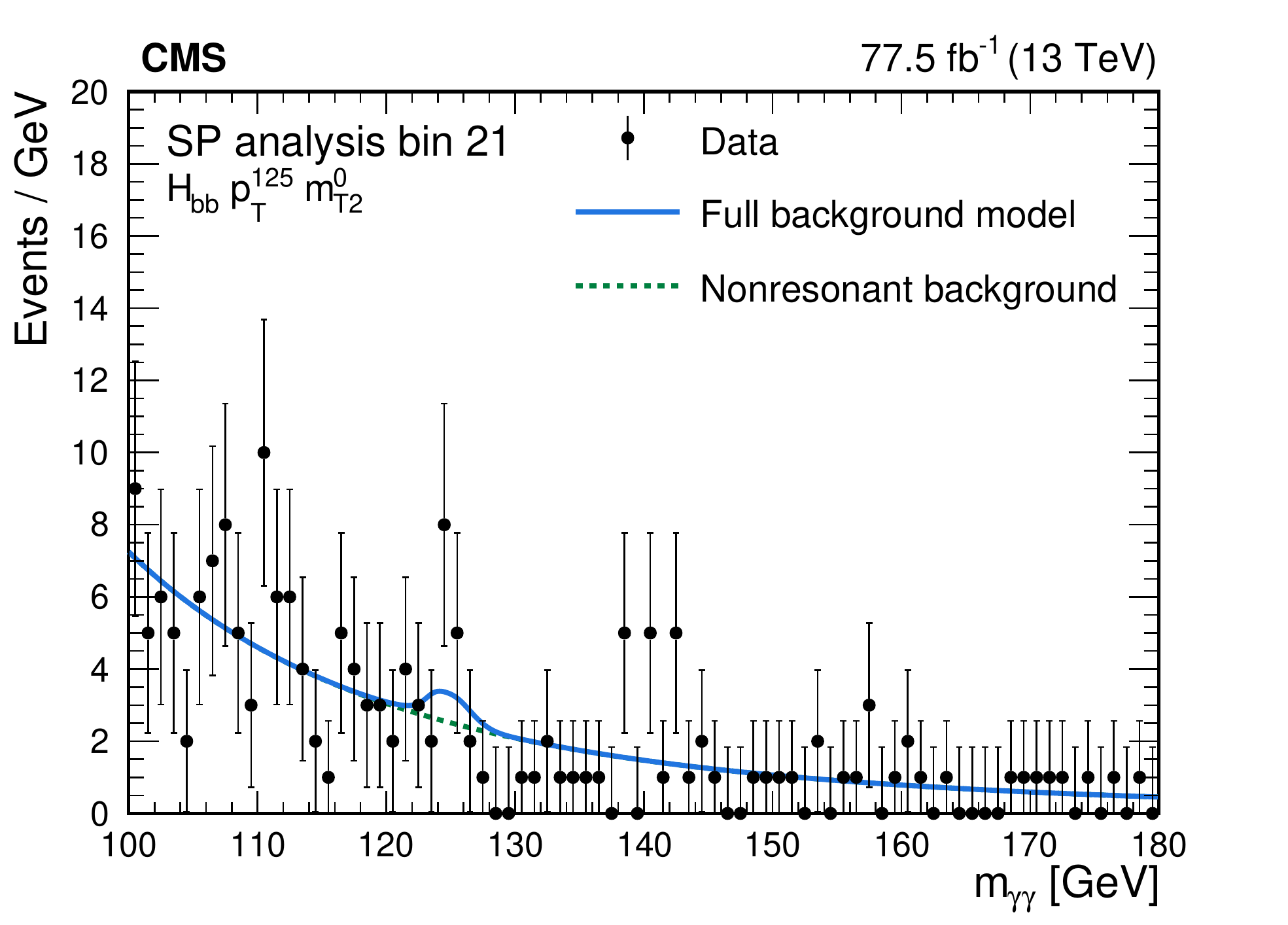}
\includegraphics[width=0.49\textwidth]{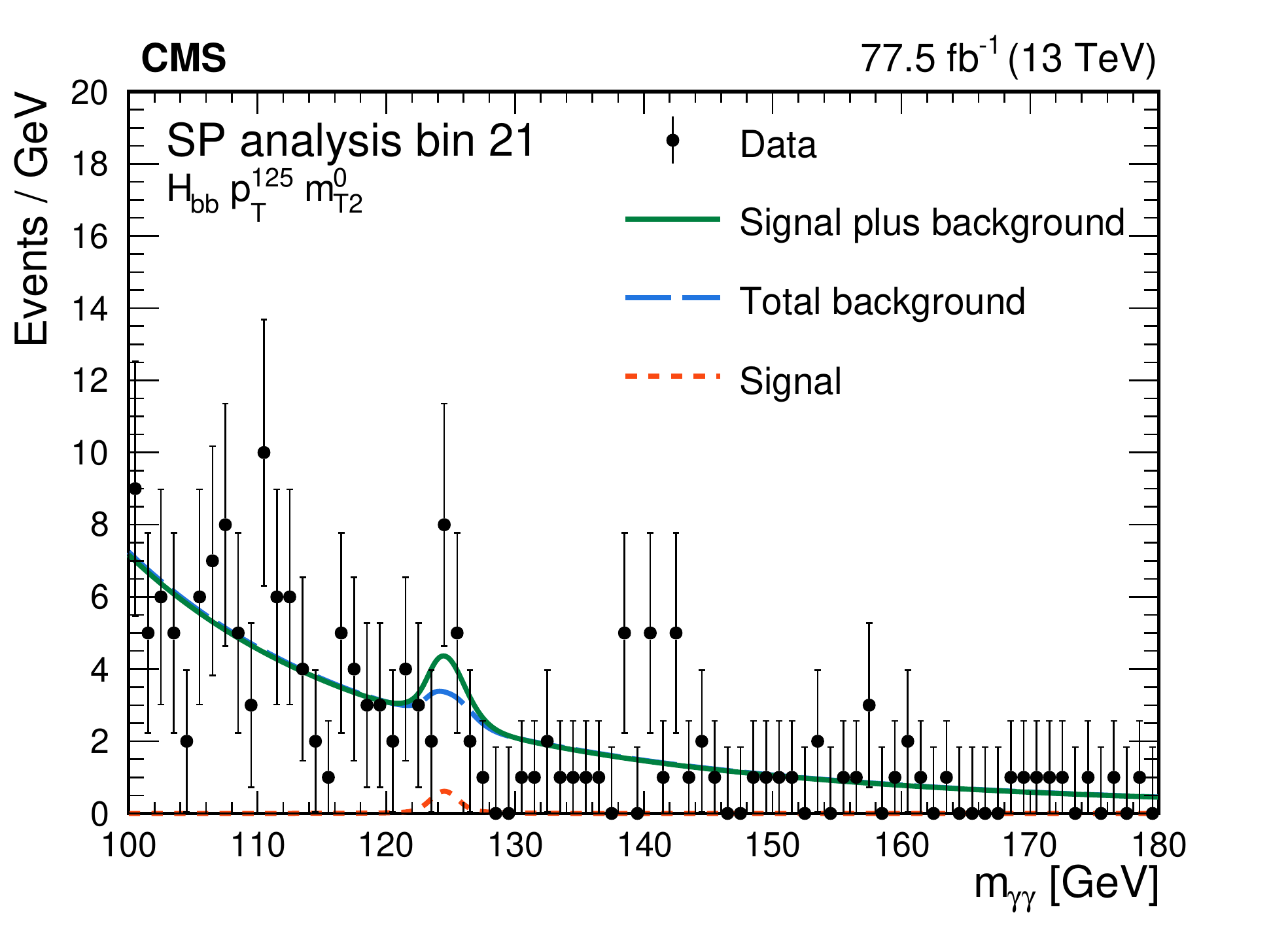}
\includegraphics[width=0.49\textwidth]{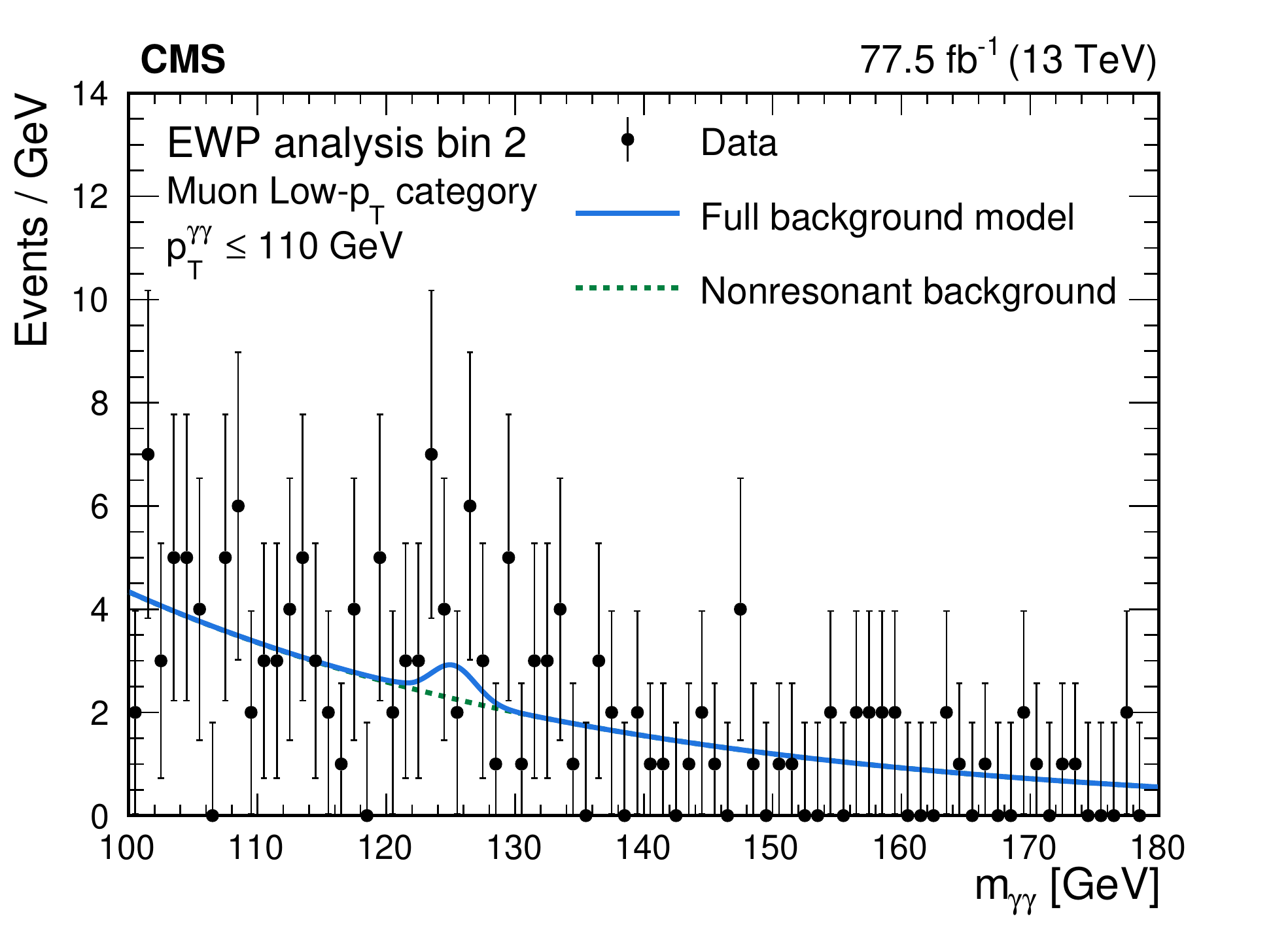}
\includegraphics[width=0.49\textwidth]{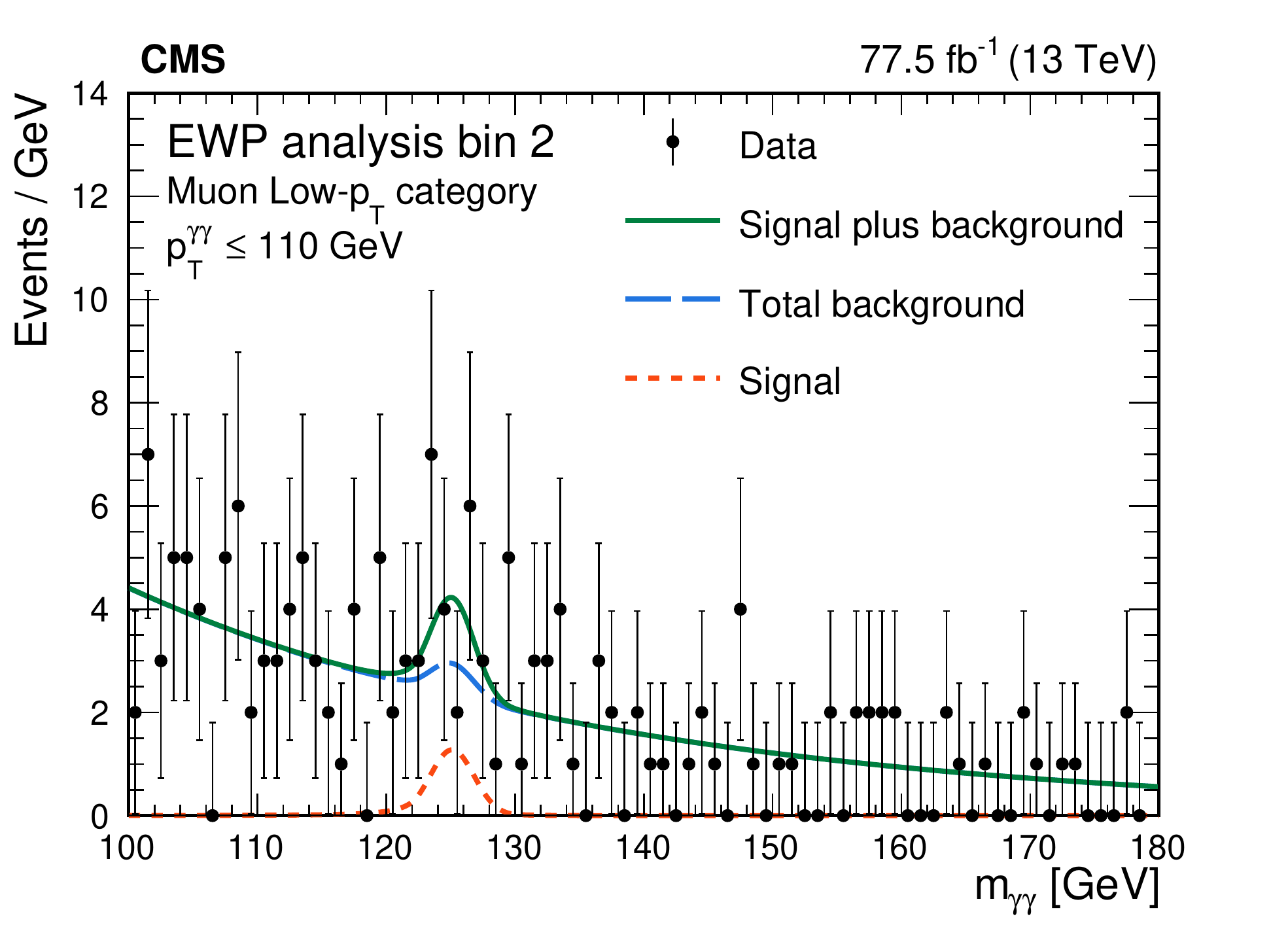}
\caption{The diphoton mass distribution for two example search bin is shown with the
background-only fit (\cmsLeft) and the signal-plus-background fit (\cmsRight)
to illustrate the signal extraction procedure.
The search region bins shown corresponds to the $\PH\PQb\cPaqb ~ \pt^{125}, ~ \mTii^{0} ~ $ category, bin 21, of the
SP analysis (upper) and the Muon Low-$\pt$ category, bin 2, of the EWP analysis (lower).
\label{fig:ExampleFit}}
\end{figure}

The fit results for the search region bins including the data yields, fitted background, and signal yields are summarized in Tables~\ref{tab:bkgYieldTableB_1}~and~\ref{tab:bkgYieldTableB_2} for the SP analysis and in Table~\ref{tab:bkgYieldTableA} for the EWP analysis. Example fit results are shown in Fig.~\ref{fig:ExampleFit} to illustrate the background-only and signal plus background fits. We observe no statistically significant deviation from the SM background expectation.

\begin{table*}[htb]
\centering
\topcaption{The observed data, fitted nonresonant background yields, and SM Higgs boson background yields within
the mass window between 122 and 129\GeV are shown for each search region bin in the $\PH\PQb\cPaqb$, $\PZ\PQb\cPaqb$, and
leptonic categories of the SP analysis. The uncertainties quoted are the fit uncertainties, which include the impact of
all systematic uncertainties. The bin names give a short-form description of the search region bin definition
which are given in full in Table~\ref{tab:SPBins1}.
The labels $\pt^{0}$, $\pt^{75}$, and $\pt^{125}$ refer to bins defined by the requirement
that \ptOverM is less than 0.6, between 0.6 and 1.0, and greater than 1.0, respectively.
The labels $\mTii^{0}$ and $\mTii^{30}$ refer to bins defined by the requirement that \mTii is
less than and greater than 30\GeV, respectively.
}

\label{tab:bkgYieldTableB_1}
\begin{tabular}{ccccc}
\hline
Search     & \multirow{2}{*}{Bin name} & Observed & Fitted           & SM Higgs boson \\
region bin &                           & data     & nonresonant bkg & bkg      \\
\hline
SP 0  & $\PZ_{\ell\ell}$                       & 2 & 1.7 $\pm$ 0.2  & 0.84 $\pm$ 0.09  \\  [\cmsTabSkip]
SP 1  & $1\PGm ~ \pt^{0}, ~ \mTii^{0} ~ $    & 24 & 20.0 $\pm$ 0.9  & 1.6 $\pm$ 0.1  \\
SP 2  & $1\PGm ~ \pt^{0}, ~ \mTii^{30} ~ $   & 10 & 8.9 $\pm$ 1.4  & 1.1 $\pm$ 0.1  \\
SP 3  & $1\PGm ~ \pt^{75}, ~ \mTii^{0} ~ $   & 3  & 2.6 $\pm$ 0.5  & 0.89 $\pm$ 0.07  \\
SP 4  & $1\PGm ~ \pt^{75}, ~ \mTii^{30} ~ $  & 7  & 2.4 $\pm$ 0.4  & 0.79 $\pm$ 0.07  \\
SP 5  & $1\PGm ~ \pt^{125}, ~ \mTii^{0} ~ $  & 4  & 3.1 $\pm$ 0.4  & 1.0 $\pm$ 0.1  \\
SP 6  & $1\PGm ~ \pt^{125}, ~ \mTii^{30} ~ $ & 3  & 2.2 $\pm$ 0.4  & 1.1 $\pm$ 0.1  \\  [\cmsTabSkip]
SP 7  & $1\Pe ~ \pt^{0}, ~ \mTii^{0} ~ $    & 93 & 87.2 $\pm$ 10.6  & 1.1 $\pm$ 0.1  \\
SP 8  & $1\Pe ~ \pt^{0}, ~ \mTii^{30} ~ $   & 15 & 13.8 $\pm$ 0.9  & 0.59 $\pm$ 0.05  \\
SP 9  & $1\Pe ~ \pt^{75}, ~ \mTii^{0} ~ $   & 10 & 18.6 $\pm$ 3.0  & 0.74 $\pm$ 0.06  \\
SP 10 & $1\Pe ~ \pt^{75}, ~ \mTii^{30} ~ $  & 3 & 4.3 $\pm$ 0.3  & 0.48 $\pm$ 0.04  \\
SP 11 & $1\Pe ~ \pt^{125}, ~ \mTii^{0} ~ $  & 7 & 6.2 $\pm$ 0.4  & 1.1 $\pm$ 0.1  \\
SP 12 & $1\Pe ~ \pt^{125}, ~ \mTii^{30} ~ $ & 1 & 1.4 $\pm$ 0.2  & 0.89 $\pm$ 0.08  \\ [\cmsTabSkip]
SP 13 & $\PZ\PQb\cPaqb ~ \pt^{0}, ~ \mTii^{0} ~ $    & 227 & 224 $\pm$ 17  & 4.4 $\pm$ 0.6  \\
SP 14 & $\PZ\PQb\cPaqb ~ \pt^{75}, ~ \mTii^{0} ~ $   & 33 & 42.2 $\pm$ 7.4  & 1.7 $\pm$ 0.2  \\
SP 15 & $\PZ\PQb\cPaqb ~ \pt^{125}, ~ \mTii^{0} ~ $  & 15 & 15.7 $\pm$ 3.6  & 2.9 $\pm$ 0.3  \\
SP 16 & $\PZ\PQb\cPaqb ~ \pt^{0}, ~ \mTii^{30} ~ $   & 44 & 43.4 $\pm$ 7.5  & 0.83 $\pm$ 0.40  \\
SP 17 & $\PZ\PQb\cPaqb ~ \pt^{75}, ~ \mTii^{30} ~ $  & 13 & 10.8 $\pm$ 2.3  & 0.48 $\pm$ 0.13  \\
SP 18 & $\PZ\PQb\cPaqb ~ \pt^{125}, ~ \mTii^{30} ~ $ & 5 & 4.5 $\pm$ 0.4  & 0.82 $\pm$ 0.11  \\ [\cmsTabSkip]
SP 19 & $\PH\PQb\cPaqb ~ \pt^{0}, ~ \mTii^{0} ~ $    & 179 & 179 $\pm$ 15  & 3.4 $\pm$ 0.3  \\
SP 20 & $\PH\PQb\cPaqb ~ \pt^{75}, ~ \mTii^{0} ~ $   & 45 & 41.2 $\pm$ 1.9  & 1.9 $\pm$ 0.2  \\
SP 21 & $\PH\PQb\cPaqb ~ \pt^{125}, ~ \mTii^{0} ~ $  & 22 & 18.4 $\pm$ 1.8  & 3.0 $\pm$ 0.9  \\
SP 22 & $\PH\PQb\cPaqb ~ \pt^{0}, ~ \mTii^{30} ~ $   & 47 & 42.5 $\pm$ 7.4  & 0.93 $\pm$ 0.32  \\
SP 23 & $\PH\PQb\cPaqb ~ \pt^{75}, ~ \mTii^{30} ~ $  & 13 & 12.1 $\pm$ 0.8  & 0.62 $\pm$ 0.06  \\
SP 24 & $\PH\PQb\cPaqb ~ \pt^{125}, ~ \mTii^{30} ~ $ & 6 & 4.4 $\pm$ 0.7  & 1.3 $\pm$ 0.2  \\
\hline
\end{tabular}
\end{table*}

\begin{table*}[htb]
\centering
\topcaption{The observed data, fitted nonresonant background yields, and SM Higgs boson background yields within the mass window
between 122 and 129\GeV are shown for each search region bin in the all-hadronic categories of the SP analysis. The uncertainties
quoted are the fit uncertainties, which include the impact of all systematic uncertainties.
The bin names give a short-form description of the search region bin definition
which are given in full in Table~\ref{tab:SPBins2}.
The labels $\pt^{0}$, $\pt^{75}$, and $\pt^{125}$ refer to bins defined by the requirement
that \ptOverM is less than 0.6, between 0.6 and 1.0, and greater than 1.0, respectively.
The labels $\mTii^{0}$ and $\mTii^{30}$ refer to bins defined by the requirement that \mTii is
less than and greater than 30\GeV, respectively.
}
\label{tab:bkgYieldTableB_2}
\begin{tabular}{ccccc}
\hline
Search     & \multirow{2}{*}{Bin name} & Observed & Fitted           & SM Higgs boson \\
region bin &                           & data     & nonresonant bkg & bkg      \\
\hline
SP 25 & $0     \text{j}, ~ \ge$0$\cPqb,  ~ \pt^{0} ~ $   & 53\,252 & 53\,662 $\pm$ 104  & 973 $\pm$ 68  \\
SP 26 & $0     \text{j}, ~ \ge$0$\cPqb,  ~ \pt^{75} ~ $  & 586 & 574 $\pm$ 27  & 33.3 $\pm$ 4.1  \\
SP 27 & $0     \text{j}, ~ \ge$0$\cPqb,  ~ \pt^{125} ~ $ & 51 & 49.5 $\pm$ 8.0  & 7.4 $\pm$ 0.8  \\  [\cmsTabSkip]
SP 28 & $1$--$3 \text{j}, ~ 0\cPqb, ~ \pt^{0}, ~ \mTii^{0} ~ $         & 14\,648 & 14\,753 $\pm$ 138  & 308 $\pm$ 33  \\
SP 29 & $1$--$3 \text{j}, ~ 0\cPqb, ~ \pt^{0}, ~ \mTii^{30} ~ $         & 2732 & 2725 $\pm$ 10  & 125 $\pm$ 10  \\
SP 30 & $1$--$3 \text{j}, ~ 0\cPqb, ~ \pt^{75}, ~ \mTii^{0} ~ $         & 781 & 708 $\pm$ 30  & 101 $\pm$ 9  \\
SP 31 & $1$--$3 \text{j}, ~ 0\cPqb, ~ \pt^{75}, ~ \mTii^{30} ~ $         & 103 & 101 $\pm$ 11  & 0.90 $\pm$ 0.38  \\
SP 32 & $1$--$3 \text{j}, ~ 0\cPqb, ~ \pt^{125}, ~ \mTii^{0} ~ $         & 47 & 46.6 $\pm$ 7.7  & 0.95 $\pm$ 0.28  \\
SP 33 & $1$--$3 \text{j}, ~ 0\cPqb, ~ \pt^{125}, ~ \mTii^{30} ~ $         & 52 & 37.2 $\pm$ 6.9  & 3.9 $\pm$ 0.6  \\
SP 34 & $1$--$3 \text{j}, ~ 1\cPqb, ~ \pt^{0}, ~ \mTii^{0} ~ $         & 4184 & 4149 $\pm$ 7  & 78.4 $\pm$ 7.7  \\
SP 35 & $1$--$3 \text{j}, ~ 1\cPqb, ~ \pt^{0}, ~ \mTii^{30} ~ $         & 928 & 902 $\pm$ 34  & 35.3 $\pm$ 3.1  \\
SP 36 & $1$--$3 \text{j}, ~ 1\cPqb, ~ \pt^{75}, ~ \mTii^{0} ~ $         & 273 & 270 $\pm$ 19  & 36.4 $\pm$ 3.1  \\
SP 37 & $1$--$3 \text{j}, ~ 1\cPqb, ~ \pt^{75}, ~ \mTii^{30} ~ $         & 75 & 78.0 $\pm$ 10.0  & 1.3 $\pm$ 0.1  \\
SP 38 & $1$--$3 \text{j}, ~ 1\cPqb, ~ \pt^{125}, ~ \mTii^{0} ~ $         & 52 & 43.7 $\pm$ 7.5  & 0.97 $\pm$ 0.26  \\
SP 39 & $1$--$3 \text{j}, ~ 1\cPqb, ~ \pt^{125}, ~ \mTii^{30} ~ $         & 38 & 30.8 $\pm$ 6.3  & 3.7 $\pm$ 0.8  \\
SP 40 & $1$--$3 \text{j}, ~ \ge$2$\cPqb, ~ \pt^{0}, ~ \mTii^{0} ~ $         & 312 & 292 $\pm$ 19  & 5.6 $\pm$ 0.8  \\
SP 41 & $1$--$3 \text{j}, ~ \ge$2$\cPqb, ~ \pt^{0}, ~ \mTii^{30} ~ $         & 79 & 79.6 $\pm$ 10.1  & 3.0 $\pm$ 0.3  \\
SP 42 & $1$--$3 \text{j}, ~ \ge$2$\cPqb, ~ \pt^{75}, ~ \mTii^{0} ~ $         & 37 & 34.3 $\pm$ 6.6  & 4.5 $\pm$ 0.6  \\
SP 43 & $1$--$3 \text{j}, ~ \ge$2$\cPqb, ~ \pt^{75}, ~ \mTii^{30} ~ $         & 26 & 24.0 $\pm$ 5.6  & 0.57 $\pm$ 0.06  \\
SP 44 & $1$--$3 \text{j}, ~ \ge$2$\cPqb, ~ \pt^{125}, ~ \mTii^{0} ~ $         & 16 & 12.3 $\pm$ 0.8  & 0.54 $\pm$ 0.10  \\
SP 45 & $1$--$3 \text{j}, ~ \ge$2$\cPqb, ~ \pt^{125}, ~ \mTii^{30} ~ $         & 15 & 10.0 $\pm$ 0.8  & 1.7 $\pm$ 0.2  \\  [\cmsTabSkip]
SP 46 & $\ge$4$\text{j}, ~ 0\cPqb, ~ \pt^{0}, ~ \mTii^{0} ~ $           & 2429 & 2426 $\pm$ 7  & 35.3 $\pm$ 2.6  \\
SP 47 & $\ge$4$\text{j}, ~ 0\cPqb, ~ \pt^{0}, ~ \mTii^{30} ~ $           & 339 & 339 $\pm$ 21  & 12.9 $\pm$ 1.2  \\
SP 48 & $\ge$4$\text{j}, ~ 0\cPqb, ~ \pt^{75}, ~ \mTii^{0} ~ $            & 118 & 97.8 $\pm$ 11.2  & 11.1 $\pm$ 2.2  \\
SP 49 & $\ge$4$\text{j}, ~ 0\cPqb, ~ \pt^{75}, ~ \mTii^{30} ~ $          & 15 & 19.5 $\pm$ 3.1  & 0.16 $\pm$ 0.05  \\
SP 50 & $\ge$4$\text{j}, ~ 0\cPqb, ~ \pt^{125}, ~ \mTii^{0} ~ $          & 13 & 10.0 $\pm$ 1.7  & 0.08 $\pm$ 1.76  \\
SP 51 & $\ge$4$\text{j}, ~ 0\cPqb, ~ \pt^{125}, ~ \mTii^{30} ~ $          & 7 & 6.5 $\pm$ 0.6  & 0.73 $\pm$ 0.18  \\
SP 52 & $\ge$4$\text{j}, ~ 1\cPqb, ~ \pt^{0}, ~ \mTii^{0} ~ $          & 833 & 800 $\pm$ 32  & 12.3 $\pm$ 2.5  \\
SP 53 & $\ge$4$\text{j}, ~ 1\cPqb, ~ \pt^{0}, ~ \mTii^{30} ~ $          & 132 & 135 $\pm$ 13  & 4.6 $\pm$ 0.3  \\
SP 54 & $\ge$4$\text{j}, ~ 1\cPqb, ~ \pt^{75}, ~ \mTii^{0} ~ $          & 33 & 42.5 $\pm$ 7.4  & 4.8 $\pm$ 0.7  \\
SP 55 & $\ge$4$\text{j}, ~ 1\cPqb, ~ \pt^{75}, ~ \mTii^{30} ~ $          & 13 & 20.2 $\pm$ 5.1  & 0.35 $\pm$ 0.04  \\
SP 56 & $\ge$4$\text{j}, ~ 1\cPqb, ~ \pt^{125}, ~ \mTii^{0} ~ $          & 10 & 11.4 $\pm$ 1.5  & 0.34 $\pm$ 0.04  \\
SP 57 & $\ge$4$\text{j}, ~ 1\cPqb, ~ \pt^{125}, ~ \mTii^{30} ~ $          & 9 & 8.4 $\pm$ 0.6  & 0.97 $\pm$ 0.11  \\
SP 58 & $\ge$4$\text{j}, ~ \ge$2$\cPqb, ~ \pt^{0}, ~ \mTii^{0} ~ $          & 90 & 88.4 $\pm$ 10.7  & 1.1 $\pm$ 0.3  \\
SP 59 & $\ge$4$\text{j}, ~ \ge$2$\cPqb, ~ \pt^{0}, ~ \mTii^{30} ~ $          & 25 & 20.9 $\pm$ 4.6  & 0.52 $\pm$ 0.06  \\
SP 60 & $\ge$4$\text{j}, ~ \ge$2$\cPqb, ~ \pt^{75}, ~ \mTii^{0} ~ $          & 11 & 8.7 $\pm$ 0.6  & 0.84 $\pm$ 0.17  \\
SP 61 & $\ge$4$\text{j}, ~ \ge$2$\cPqb, ~ \pt^{75}, ~ \mTii^{30} ~ $          & 12 & 11.5 $\pm$ 3.7  & 0.26 $\pm$ 0.09  \\
SP 62 & $\ge$4$\text{j}, ~ \ge$2$\cPqb, ~ \pt^{125}, ~ \mTii^{0} ~ $          & 6 & 3.7 $\pm$ 0.4  & 0.24 $\pm$ 0.08  \\
SP 63 & $\ge$4$\text{j}, ~ \ge$2$\cPqb, ~ \pt^{125}, ~ \mTii^{30} ~ $         & 4 & 5.2 $\pm$ 1.1  & 0.69 $\pm$ 0.09  \\
\hline
\end{tabular}
\end{table*}

\begin{table*}[htb]
\centering
\topcaption{The observed data, fitted nonresonant background yields, and SM Higgs boson
background yields within the mass window between 122 and 129\GeV are shown for each
search region bin of the EWP analysis. The uncertainties quoted are the fit
uncertainties, which include the impact of all systematic uncertainties.
}

\begin{tabular}{ccccc}
\hline
Search     & \multirow{2}{*}{Category} & Observed & Fitted           & SM Higgs boson \\
region bin &                           & data     & nonresonant bkg & bkg      \\
\hline

EWP 0 & Two-Lepton                           &         2 &     1.5 $\pm$  0.4  &    1.1 $\pm$ 0.6 \\
EWP 1 & Muon High-$\pt$                      &        11 &     6.2 $\pm$  0.9  &    3.7 $\pm$ 0.8 \\
EWP 2 & Muon Low-$\pt$                       &        28 &    15.8 $\pm$  1.4  &    3.0 $\pm$ 0.8 \\
EWP 3 & Electron High-$\pt$                  &        17 &    11.9 $\pm$  1.3  &    3.4 $\pm$ 1.1 \\
EWP 4 & Electron Low-$\pt$                   &         8 &     5.2 $\pm$  0.8  &    0.6 $\pm$ 0.2 \\
EWP 5 & Electron Low-$\pt$                   &        18 &    31.5 $\pm$  1.9  &    0.9 $\pm$ 0.4 \\
EWP 6 & Electron Low-$\pt$                   &         9 &    13.7 $\pm$  1.3  &    0.7 $\pm$ 0.3 \\
EWP 7 & $\PH\PQb\cPaqb$ High-$\pt$         &         9 &     7.0 $\pm$  0.9  &    1.2 $\pm$ 0.4 \\
EWP 8 & $\PH\PQb\cPaqb$ High-$\pt$         &        19 &    17.8 $\pm$  1.5  &    3.8 $\pm$ 0.7 \\
EWP 9 & $\PH\PQb\cPaqb$ Low-$\pt$          &        34 &    25.8 $\pm$  1.8  &    0.8 $\pm$ 0.1 \\
EWP 10 & $\PH\PQb\cPaqb$ Low-$\pt$         &        60 &    51.0 $\pm$  2.4  &    1.9 $\pm$ 0.3 \\
EWP 11 & $\PZ\PQb\cPaqb$ High-$\pt$        &         3 &     7.2 $\pm$  1.1  &    0.5 $\pm$ 0.1 \\
EWP 12 & $\PZ\PQb\cPaqb$ High-$\pt$        &        17 &    14.0 $\pm$  1.3  &    2.8 $\pm$ 1.1 \\
EWP 13 & $\PZ\PQb\cPaqb$ High-$\pt$        &        10 &     9.4 $\pm$  1.1  &    1.3 $\pm$ 0.3 \\
EWP 14 & $\PZ\PQb\cPaqb$ Low-$\pt$         &        27 &    35.2 $\pm$  2.0  &    0.8 $\pm$ 0.2 \\
EWP 15 & $\PZ\PQb\cPaqb$ Low-$\pt$         &        84 &    75.1 $\pm$  2.9  &    2.5 $\pm$ 1.3 \\
EWP 16 & $\PZ\PQb\cPaqb$ Low-$\pt$         &        45 &    46.3 $\pm$  2.3  &    1.2 $\pm$ 0.4 \\
EWP 17 & High-$\pt$                          &        11 &    14.4 $\pm$  1.3  &    1.8 $\pm$ 0.2 \\
EWP 18 & High-$\pt$                          &        31 &    21.8 $\pm$  1.6  &    2.1 $\pm$ 0.4 \\
EWP 19 & High-$\pt$                          &        11 &    13.5 $\pm$  1.3  &    1.2 $\pm$ 0.3 \\
EWP 20 & High-$\pt$                          &      1834 &    1648 $\pm$   14  &    248 $\pm$  38 \\
EWP 21 & High-$\pt$                          &        91 &   100.2 $\pm$  3.7  &    8.9 $\pm$ 1.5 \\
EWP 22 & High-$\pt$                          &        12 &    14.4 $\pm$  1.4  &    1.2 $\pm$ 0.2 \\
EWP 23 & High-Res                            &        30 &    20.6 $\pm$  1.6  &    0.6 $\pm$ 0.2 \\
EWP 24 & High-Res                            &        46 &    49.1 $\pm$  4.0  &    1.5 $\pm$ 0.5 \\
EWP 25 & High-Res                            &         9 &    17.0 $\pm$  1.4  &    0.4 $\pm$ 0.1 \\
EWP 26 & High-Res                            &      5186 &    5057 $\pm$   25  &    219 $\pm$  42 \\
EWP 27 & High-Res                            &        53 &    63.0 $\pm$  2.6  &    2.4 $\pm$ 1.0 \\
EWP 28 & High-Res                            &        19 &    17.7 $\pm$  1.5  &    0.5 $\pm$ 0.1 \\
EWP 29 & Low-Res                             &        26 &    33.8 $\pm$  2.1  &    0.3 $\pm$ 0.1 \\
EWP 30 & Low-Res                             &        61 &    65.8 $\pm$  3.0  &    0.9 $\pm$ 0.2 \\
EWP 31 & Low-Res                             &        24 &    18.3 $\pm$  1.5  &    0.2 $\pm$ 0.1 \\
EWP 32 & Low-Res                             &      5548 &    5328 $\pm$   22  &    141 $\pm$  27 \\
EWP 33 & Low-Res                             &        78 &    79.1 $\pm$  2.9  &    1.4 $\pm$ 0.4 \\
EWP 34 & Low-Res                             &        25 &    23.7 $\pm$  1.8  &    0.4 $\pm$ 0.1 \\
\hline
\end{tabular}
\label{tab:bkgYieldTableA}
\end{table*}

The search results are interpreted in terms of limits on the product of the production cross section and
branching fraction for simplified models of bottom squark pair production and chargino-neutralino
production indicated in Fig.~\ref{fig:SMSDiagrams}. In the case of bottom squark pair production,
we consider the scenario where the bottom squark subsequently decays to a bottom quark and 
the next-to-lightest neutralino ($\PSGczDt$), where the $\PSGczDt$ decays
to a Higgs boson and the LSP ($\PSGczDo$). The mass splitting between the $\PSGczDt$ and $\PSGczDo$
is assumed to be 130\GeV, slightly above threshold to produce an on-shell Higgs boson.

In the case of chargino-neutralino production, we consider two different scenarios. In the first scenario,
the pure wino-like charginos ($\PSGcpmDo$) and the $\PSGczDt$ are mass-degenerate and are produced together,
with the chargino decaying to a $\PW$ boson and the $\PSGczDo$ LSP, and the $\PSGczDt$ decaying to a Higgs
boson and the LSP. The production cross sections are computed at NLO+NLL accuracy in QCD in the limit of
mass-degenerate wino $\PSGczDt$ and $\PSGcpmDo$, light bino $\PSGczDo$, and with all the other sparticles
assumed to be heavy and decoupled~\cite{Beenakker:1999xh,Fuks:2012qx,Fuks:2013vua}.

In the second scenario, we consider a GMSB~\cite{Dimopoulos:1996vz,Matchev:1999ft}
simplified model where higgsino-like charg\-inos and neutralinos are nearly
mass-degenerate and are produced in pairs through the following
combinations: $\PSGczDo\PSGczDt$, $\PSGczDo\PSGcpmDo$, $\PSGczDt\PSGcpmDo$,
and $\PSGcpmDo\PSGcmpDo$. Because of the mass degeneracy, both the $\PSGczDt$
and $\PSGcpmDo$ will decay to $\PSGczDo$ and other low-$\pt$ (soft) particles,
leading to a signature with a $\PSGczDo$ pair. Each $\PSGczDo$ will subsequently
decay to a Higgs boson and the $\PXXSG$ LSP, or to a \PZ boson and the LSP. We
consider the case where the branching fraction of the $\PSGczDo\to \PH\PXXSG$
decay is 100\%, and the case where the branching fraction of
the $\PSGczDo\to \PH\PXXSG$ and $\PSGczDo\to \PZ\PXXSG$
decays are each $50\%$. This scenario is represented by the
$\PSGczDo$-pair production simplified model shown on Fig.~\ref{fig:SMSDiagrams}.

We show the expected event yields from a representative selection of the
different simplified SUSY models considered in the different
search region bins of the SP analysis in Tables~\ref{tab:signalYieldTableB_1}~and~\ref{tab:signalYieldTableB_2},
and in the different search region bins of the EWP analysis in Table~\ref{tab:signalYieldTableA}.
The details of the particular signal model are described in the caption of Table~\ref{tab:signalYieldTableB_1}.

\begin{table*}[htb]
\centering
\topcaption{The expected signal yields for the SUSY simplified model signals considered are shown for
each search region bin in the $\PH\PQb\cPaqb$, $\PZ\PQb\cPaqb$, and leptonic categories of the SP analysis.
The bin names give a short-form description of the search region bin definition
which are given in full in Table~\ref{tab:SPBins1}.
The labels $\pt^{0}$, $\pt^{75}$, and $\pt^{125}$ refer to bins defined by the requirement
that \ptOverM is less than 0.6, between 0.6 and 1.0, and greater than 1.0, respectively.
The labels $\mTii^{0}$ and $\mTii^{30}$ refer to bins defined by the requirement that \mTii is
less than and greater than 30\GeV, respectively.
The labels $\PH\PH$ and $\PZ\PH$ refer to the signal models for higgsino-like
chargino and neutralino production where the
branching fractions of the decays $\PSGczDo\to \PH\PXXSG$ and $\PSGczDo\to \PZ\PXXSG$ are $100\%$ and $0\%$
, and $50\%$ and $50\%$, respectively. For the above two scenarios, the mass of the chargino and next-to-lightest
neutralino is 175\GeV, while the LSP mass is 45\GeV.
The label $\PW\PH$ (200,1) refers to the signal model for wino-like
chargino and neutralino production, where the mass of the chargino and next-to-lightest
neutralino is 200\GeV and the LSP mass is 1\GeV.
The labels \PSQb(450,1) and \PSQb(450,300) refer to
the signal models for bottom squark pair production where the bottom squark mass is 450\GeV
and the LSP mass is 1~and~300\GeV, respectively.
}

\label{tab:signalYieldTableB_1}
\cmsTable{
\begin{tabular}{ccccccc}
\hline
Search     & \multirow{2}{*}{Bin name} & \multirow{2}{*}{$\PH\PH$} & \multirow{2}{*}{$\PZ\PH$}  & \multirow{2}{*}{$\PW\PH$ (200,1)} & \multirow{2}{*}{\PSQb (450,1)} & \multirow{2}{*}{\PSQb (450,300)}  \\
region bin &                           & & & & &    \\
\hline
SP 0  & $\PZ_{\ell\ell}$ & 0.15 $\pm$ 0.02        & 1.2 $\pm$ 0.2  & 0.0 $\pm$ 0.0  & 0.07 $\pm$ 0.01  & 0.10 $\pm$ 0.01  \\ [\cmsTabSkip]
SP 1  & $1\PGm ~ \pt^{0}, ~ \mTii^{0} ~ $    & 0.67 $\pm$ 0.11  & 0.22 $\pm$ 0.04  & 0.63 $\pm$ 0.07  & 0.69 $\pm$ 0.06  & 0.10 $\pm$ 0.01  \\
SP 2  & $1\PGm ~ \pt^{0}, ~ \mTii^{30} ~ $    & 0.59 $\pm$ 0.10  & 0.23 $\pm$ 0.04  & 1.1 $\pm$ 0.1  & 0.88 $\pm$ 0.07  & 0.09 $\pm$ 0.01  \\
SP 3  & $1\PGm ~ \pt^{75}, ~ \mTii^{0} ~ $    & 0.68 $\pm$ 0.09  & 0.22 $\pm$ 0.03  & 0.44 $\pm$ 0.04  & 0.40 $\pm$ 0.03  & 0.17 $\pm$ 0.01  \\
SP 4  & $1\PGm ~ \pt^{75}, ~ \mTii^{30} ~ $    & 0.74 $\pm$ 0.09  & 0.27 $\pm$ 0.03  & 1.0 $\pm$ 0.1  & 0.45 $\pm$ 0.04  & 0.18 $\pm$ 0.01  \\
SP 5  & $1\PGm ~ \pt^{125}, ~ \mTii^{0} ~ $    & 1.6 $\pm$ 0.3  & 0.51 $\pm$ 0.08  & 0.72 $\pm$ 0.14  & 0.24 $\pm$ 0.02  & 1.2 $\pm$ 0.1  \\
SP 6  & $1\PGm ~ \pt^{125}, ~ \mTii^{30} ~ $    & 1.7 $\pm$ 0.3  & 0.58 $\pm$ 0.10  & 1.7 $\pm$ 0.3  & 0.32 $\pm$ 0.03  & 1.6 $\pm$ 0.1  \\  [\cmsTabSkip]
SP 7  & $1\Pe ~ \pt^{0}, ~ \mTii^{0} ~ $    & 0.43 $\pm$ 0.12  & 0.18 $\pm$ 0.03  & 0.41 $\pm$ 0.05  & 0.52 $\pm$ 0.04  & 0.06 $\pm$ 0.00  \\
SP 8  & $1\Pe ~ \pt^{0}, ~ \mTii^{30} ~ $   & 0.43 $\pm$ 0.11  & 0.19 $\pm$ 0.04  & 0.78 $\pm$ 0.12  & 0.52 $\pm$ 0.03  & 0.05 $\pm$ 0.00  \\
SP 9  & $1\Pe ~ \pt^{75}, ~ \mTii^{0} ~ $   & 0.45 $\pm$ 0.11  & 0.19 $\pm$ 0.02  & 0.30 $\pm$ 0.03  & 0.27 $\pm$ 0.02  & 0.12 $\pm$ 0.01  \\
SP 10 & $1\Pe ~ \pt^{75}, ~ \mTii^{30} ~ $   & 0.48 $\pm$ 0.09  & 0.22 $\pm$ 0.02  & 0.66 $\pm$ 0.07  & 0.29 $\pm$ 0.02  & 0.12 $\pm$ 0.01  \\
SP 11 & $1\Pe ~ \pt^{125}, ~ \mTii^{0} ~ $   & 1.3 $\pm$ 0.3  & 0.46 $\pm$ 0.09  & 0.60 $\pm$ 0.11  & 0.24 $\pm$ 0.02  & 0.87 $\pm$ 0.07  \\
SP 12 & $1\Pe ~ \pt^{125}, ~ \mTii^{30} ~ $   & 1.5 $\pm$ 0.3  & 0.57 $\pm$ 0.09  & 1.4 $\pm$ 0.3  & 0.28 $\pm$ 0.02  & 1.1 $\pm$ 0.1  \\ [\cmsTabSkip]
SP 13 & $\PZ\PQb\cPaqb ~ \pt^{0}, ~ \mTii^{0} ~ $       & 1.3 $\pm$ 0.2  & 0.50 $\pm$ 0.08  & 0.09 $\pm$ 0.02  & 3.0 $\pm$ 0.2  & 0.29 $\pm$ 0.02  \\
SP 14 & $\PZ\PQb\cPaqb ~ \pt^{75}, ~ \mTii^{0} ~ $    & 1.3 $\pm$ 0.1  & 0.52 $\pm$ 0.06  & 0.05 $\pm$ 0.01  & 1.7 $\pm$ 0.1  & 0.63 $\pm$ 0.04  \\
SP 15 & $\PZ\PQb\cPaqb ~ \pt^{125}, ~ \mTii^{0} ~ $    & 2.9 $\pm$ 0.5  & 1.2 $\pm$ 0.2  & 0.11 $\pm$ 0.02  & 1.3 $\pm$ 0.1  & 5.1 $\pm$ 0.3  \\
SP 16 & $\PZ\PQb\cPaqb ~ \pt^{0}, ~ \mTii^{30} ~ $    & 1.1 $\pm$ 0.2  & 0.49 $\pm$ 0.08  & 0.12 $\pm$ 0.02  & 2.5 $\pm$ 0.3  & 0.13 $\pm$ 0.01  \\
SP 17 & $\PZ\PQb\cPaqb ~ \pt^{75}, ~ \mTii^{30} ~ $    & 1.1 $\pm$ 0.1  & 0.52 $\pm$ 0.07  & 0.13 $\pm$ 0.02  & 1.5 $\pm$ 0.1  & 0.31 $\pm$ 0.03  \\
SP 18 & $\PZ\PQb\cPaqb ~ \pt^{125}, ~ \mTii^{30} ~ $    & 2.3 $\pm$ 0.4  & 1.3 $\pm$ 0.2  & 0.25 $\pm$ 0.05  & 1.1 $\pm$ 0.1  & 2.2 $\pm$ 0.2  \\ [\cmsTabSkip]
SP 19 & $\PH\PQb\cPaqb ~ \pt^{0}, ~ \mTii^{0} ~ $       & 2.9 $\pm$ 0.5  & 0.81 $\pm$ 0.14  & 0.03 $\pm$ 0.01  & 5.9 $\pm$ 0.4  & 1.4 $\pm$ 0.1  \\
SP 20 & $\PH\PQb\cPaqb ~ \pt^{75}, ~ \mTii^{0} ~ $    & 3.3 $\pm$ 0.3  & 0.91 $\pm$ 0.13  & 0.04 $\pm$ 0.01  & 3.4 $\pm$ 0.3  & 2.6 $\pm$ 0.2  \\
SP 21 & $\PH\PQb\cPaqb ~ \pt^{125}, ~ \mTii^{0} ~ $    & 9.6 $\pm$ 1.8  & 2.6 $\pm$ 0.5  & 0.06 $\pm$ 0.01  & 3.0 $\pm$ 0.2  & 22.7 $\pm$ 1.7  \\
SP 22 & $\PH\PQb\cPaqb ~ \pt^{0}, ~ \mTii^{30} ~ $      & 2.5 $\pm$ 0.4  & 0.71 $\pm$ 0.10  & 0.10 $\pm$ 0.01  & 4.7 $\pm$ 0.5  & 0.49 $\pm$ 0.05  \\
SP 23 & $\PH\PQb\cPaqb ~ \pt^{75}, ~ \mTii^{30} ~ $   & 2.9 $\pm$ 0.3  & 0.82 $\pm$ 0.10  & 0.11 $\pm$ 0.02  & 3.0 $\pm$ 0.3  & 0.86 $\pm$ 0.08  \\
SP 24 & $\PH\PQb\cPaqb ~ \pt^{125}, ~ \mTii^{30} ~ $    & 8.2 $\pm$ 1.6  & 2.4 $\pm$ 0.4  & 0.15 $\pm$ 0.04  & 2.8 $\pm$ 0.2  & 8.7 $\pm$ 0.7  \\
\hline
\end{tabular}
}
\end{table*}

\begin{table*}[htb]
\centering
\topcaption{The expected signal yields for the SUSY simplified model signals considered are shown for each search region bin in the all-hadronic categories of the SP analysis.
The bin names give a short-form description of the search region bin definition
which are given in full in Table~\ref{tab:SPBins2}.
The labels $\pt^{0}$, $\pt^{75}$, and $\pt^{125}$ refer to bins defined by the requirement
that \ptOverM is less than 0.6, between 0.6 and 1.0, and greater than 1.0, respectively.
The labels $\mTii^{0}$ and $\mTii^{30}$ refer to bins defined by the requirement that \mTii is
less than and greater than 30\GeV, respectively.
The labels for the different signal models are explained in detail in the caption of Table~\ref{tab:signalYieldTableB_1}.
}
\label{tab:signalYieldTableB_2}
\cmsTable{
\begin{tabular}{ccccccc}
\hline
Search     & \multirow{2}{*}{Bin name} & \multirow{2}{*}{$\PH\PH$} & \multirow{2}{*}{$\PZ\PH$}  & \multirow{2}{*}{$\PW\PH$ (200,1)} & \multirow{2}{*}{\PSQb (450,1)} & \multirow{2}{*}{\PSQb (450,300)}  \\
region bin &                           & & & & &    \\
\hline
SP 25 & $0	\text{j}, ~ \ge$0$\cPqb,  ~ \pt^{0} ~ $                  & 3.9 $\pm$ 0.6  & 2.9 $\pm$ 0.5  & 2.6 $\pm$ 0.3  & 2.7 $\pm$ 0.1  & 0.0 $\pm$ 0.0  \\
SP 26 & $0	\text{j}, ~ \ge$0$\cPqb,  ~ \pt^{75} ~ $                 & 2.4 $\pm$ 0.3  & 2.1 $\pm$ 0.2  & 1.8 $\pm$ 0.2  & 0.54 $\pm$ 0.02  & 0.0 $\pm$ 0.0  \\
SP 27 & $0	\text{j}, ~ \ge$0$\cPqb,  ~ \pt^{125} ~ $                 & 1.7 $\pm$ 0.2  & 2.7 $\pm$ 0.4  & 1.7 $\pm$ 0.2  & 0.15 $\pm$ 0.01  & 0.01 $\pm$ 0.00  \\  [\cmsTabSkip]
SP 28 & $1$--$3	\text{j}, ~ 0\cPqb, ~ \pt^{0}, ~ \mTii^{0} ~ $                    & 4.7 $\pm$ 0.8  & 2.7 $\pm$ 0.4  & 2.9 $\pm$ 0.3  & 4.2 $\pm$ 0.5  & 0.03 $\pm$ 0.00  \\
SP 29 & $1$--$3	\text{j}, ~ 0\cPqb, ~ \pt^{0}, ~ \mTii^{30} ~ $                    & 4.7 $\pm$ 0.5  & 2.6 $\pm$ 0.3  & 2.1 $\pm$ 0.2  & 1.6 $\pm$ 0.3  & 0.03 $\pm$ 0.01  \\
SP 30 & $1$--$3	\text{j}, ~ 0\cPqb, ~ \pt^{75}, ~ \mTii^{0} ~ $                    & 9.0 $\pm$ 1.5  & 5.1 $\pm$ 0.9  & 3.1 $\pm$ 0.6  & 0.73 $\pm$ 0.15  & 0.27 $\pm$ 0.05  \\
SP 31 & $1$--$3	\text{j}, ~ 0\cPqb, ~ \pt^{75}, ~ \mTii^{30} ~ $                    & 0.21 $\pm$ 0.04  & 0.10 $\pm$ 0.02  & 0.10 $\pm$ 0.01  & 0.34 $\pm$ 0.09  & 0.04 $\pm$ 0.01  \\
SP 32 & $1$--$3	\text{j}, ~ 0\cPqb, ~ \pt^{125}, ~ \mTii^{0} ~ $                    & 0.18 $\pm$ 0.02  & 0.10 $\pm$ 0.01  & 0.07 $\pm$ 0.01  & 0.15 $\pm$ 0.04  & 0.05 $\pm$ 0.01  \\
SP 33 & $1$--$3	\text{j}, ~ 0\cPqb, ~ \pt^{125}, ~ \mTii^{30} ~ $                    & 0.66 $\pm$ 0.14  & 0.35 $\pm$ 0.07  & 0.19 $\pm$ 0.04  & 0.14 $\pm$ 0.03  & 0.35 $\pm$ 0.07  \\
SP 34 & $1$--$3	\text{j}, ~ 1\cPqb, ~ \pt^{0}, ~ \mTii^{0} ~ $                    & 6.1 $\pm$ 0.9  & 2.2 $\pm$ 0.3  & 1.1 $\pm$ 0.1  & 7.1 $\pm$ 1.0  & 0.12 $\pm$ 0.02  \\
SP 35 & $1$--$3	\text{j}, ~ 1\cPqb, ~ \pt^{0}, ~ \mTii^{30} ~ $                    & 6.6 $\pm$ 0.6  & 2.4 $\pm$ 0.2  & 0.81 $\pm$ 0.06  & 3.4 $\pm$ 0.3  & 0.20 $\pm$ 0.02  \\
SP 36 & $1$--$3	\text{j}, ~ 1\cPqb, ~ \pt^{75}, ~ \mTii^{0} ~ $                    & 13.7 $\pm$ 2.1  & 5.1 $\pm$ 0.9  & 1.4 $\pm$ 0.2  & 2.2 $\pm$ 0.3  & 1.7 $\pm$ 0.2  \\
SP 37 & $1$--$3	\text{j}, ~ 1\cPqb, ~ \pt^{75}, ~ \mTii^{30} ~ $                    & 0.23 $\pm$ 0.03  & 0.09 $\pm$ 0.01  & 0.08 $\pm$ 0.01  & 0.82 $\pm$ 0.13  & 0.27 $\pm$ 0.04  \\
SP 38 & $1$--$3	\text{j}, ~ 1\cPqb, ~ \pt^{125}, ~ \mTii^{0} ~ $                    & 0.36 $\pm$ 0.04  & 0.13 $\pm$ 0.01  & 0.07 $\pm$ 0.00  & 0.39 $\pm$ 0.06  & 0.59 $\pm$ 0.08  \\
SP 39 & $1$--$3	\text{j}, ~ 1\cPqb, ~ \pt^{125}, ~ \mTii^{30} ~ $                    & 1.2 $\pm$ 0.2  & 0.47 $\pm$ 0.09  & 0.18 $\pm$ 0.03  & 0.37 $\pm$ 0.05  & 3.5 $\pm$ 0.5  \\
SP 40 & $1$--$3	\text{j}, ~ \ge$2$\cPqb, ~ \pt^{0}, ~ \mTii^{0} ~ $                & 0.60 $\pm$ 0.09  & 0.21 $\pm$ 0.04  & 0.08 $\pm$ 0.01  & 1.9 $\pm$ 0.2  & 0.43 $\pm$ 0.05  \\
SP 41 & $1$--$3	\text{j}, ~ \ge$2$\cPqb, ~ \pt^{0}, ~ \mTii^{30} ~ $                & 0.81 $\pm$ 0.07  & 0.27 $\pm$ 0.02  & 0.07 $\pm$ 0.01  & 1.2 $\pm$ 0.1  & 0.69 $\pm$ 0.07  \\
SP 42 & $1$--$3	\text{j}, ~ \ge$2$\cPqb, ~ \pt^{75}, ~ \mTii^{0} ~ $                & 2.0 $\pm$ 0.4  & 0.67 $\pm$ 0.11  & 0.09 $\pm$ 0.03  & 0.98 $\pm$ 0.12  & 5.0 $\pm$ 0.6  \\
SP 43 & $1$--$3	\text{j}, ~ \ge$2$\cPqb, ~ \pt^{75}, ~ \mTii^{30} ~ $                & 0.08 $\pm$ 0.01  & 0.03 $\pm$ 0.01  & 0.02 $\pm$ 0.01  & 0.38 $\pm$ 0.04  & 1.3 $\pm$ 0.1  \\
SP 44 & $1$--$3	\text{j}, ~ \ge$2$\cPqb, ~ \pt^{125}, ~ \mTii^{0} ~ $                & 0.11 $\pm$ 0.03  & 0.04 $\pm$ 0.00  & 0.03 $\pm$ 0.00  & 0.28 $\pm$ 0.03  & 2.2 $\pm$ 0.2  \\
SP 45 & $1$--$3	\text{j}, ~ \ge$2$\cPqb, ~ \pt^{125}, ~ \mTii^{30} ~ $                & 0.44 $\pm$ 0.10  & 0.16 $\pm$ 0.03  & 0.05 $\pm$ 0.03  & 0.37 $\pm$ 0.03  & 15.5 $\pm$ 1.3  \\  [\cmsTabSkip]
SP 46 & $\ge$4$\text{j}, ~ 0\cPqb, ~ \pt^{0}, ~ \mTii^{0} ~ $                & 3.9 $\pm$ 0.6  & 3.1 $\pm$ 0.5  & 6.6 $\pm$ 0.7  & 3.3 $\pm$ 0.8  & 0.01 $\pm$ 0.00  \\
SP 47 & $\ge$4$\text{j}, ~ 0\cPqb, ~ \pt^{0}, ~ \mTii^{30} ~ $                & 4.2 $\pm$ 0.5  & 3.4 $\pm$ 0.4  & 5.6 $\pm$ 0.5  & 1.2 $\pm$ 0.2  & 0.03 $\pm$ 0.01  \\
SP 48 & $\ge$4$\text{j}, ~ 0\cPqb, ~ \pt^{75}, ~ \mTii^{0} ~ $                & 7.5 $\pm$ 1.2  & 6.9 $\pm$ 1.2  & 8.0 $\pm$ 1.4  & 0.56 $\pm$ 0.11  & 0.13 $\pm$ 0.03  \\
SP 49 & $\ge$4$\text{j}, ~ 0\cPqb, ~ \pt^{75}, ~ \mTii^{30} ~ $                & 0.14 $\pm$ 0.02  & 0.10 $\pm$ 0.01  & 0.19 $\pm$ 0.02  & 0.52 $\pm$ 0.11  & 0.02 $\pm$ 0.00  \\
SP 50 & $\ge$4$\text{j}, ~ 0\cPqb, ~ \pt^{125}, ~ \mTii^{0} ~ $                & 0.16 $\pm$ 0.02  & 0.13 $\pm$ 0.02  & 0.19 $\pm$ 0.02  & 0.25 $\pm$ 0.05  & 0.02 $\pm$ 0.00  \\
SP 51 & $\ge$4$\text{j}, ~ 0\cPqb, ~ \pt^{125}, ~ \mTii^{30} ~ $                & 0.81 $\pm$ 0.18  & 0.50 $\pm$ 0.11  & 0.51 $\pm$ 0.11  & 0.27 $\pm$ 0.05  & 0.16 $\pm$ 0.03  \\
SP 52 & $\ge$4$\text{j}, ~ 1\cPqb, ~ \pt^{0}, ~ \mTii^{0} ~ $                 & 5.0 $\pm$ 0.8  & 2.3 $\pm$ 0.3  & 2.5 $\pm$ 0.3  & 5.1 $\pm$ 0.9  & 0.08 $\pm$ 0.01  \\
SP 53 & $\ge$4$\text{j}, ~ 1\cPqb, ~ \pt^{0}, ~ \mTii^{30} ~ $                 & 5.4 $\pm$ 0.6  & 2.5 $\pm$ 0.2  & 2.1 $\pm$ 0.2  & 2.3 $\pm$ 0.2  & 0.15 $\pm$ 0.02  \\
SP 54 & $\ge$4$\text{j}, ~ 1\cPqb, ~ \pt^{75}, ~ \mTii^{0} ~ $                 & 11.4 $\pm$ 1.8  & 5.5 $\pm$ 0.9  & 3.5 $\pm$ 0.6  & 1.4 $\pm$ 0.2  & 1.1 $\pm$ 0.1  \\
SP 55 & $\ge$4$\text{j}, ~ 1\cPqb, ~ \pt^{75}, ~ \mTii^{30} ~ $                 & 0.27 $\pm$ 0.03  & 0.14 $\pm$ 0.02  & 0.18 $\pm$ 0.02  & 1.2 $\pm$ 0.2  & 0.11 $\pm$ 0.01  \\
SP 56 & $\ge$4$\text{j}, ~ 1\cPqb, ~ \pt^{125}, ~ \mTii^{0} ~ $                 & 0.33 $\pm$ 0.03  & 0.14 $\pm$ 0.01  & 0.17 $\pm$ 0.01  & 0.81 $\pm$ 0.13  & 0.15 $\pm$ 0.03  \\
SP 57 & $\ge$4$\text{j}, ~ 1\cPqb, ~ \pt^{125}, ~ \mTii^{30} ~ $                 & 1.4 $\pm$ 0.3  & 0.65 $\pm$ 0.12  & 0.42 $\pm$ 0.09  & 0.76 $\pm$ 0.12  & 1.5 $\pm$ 0.2  \\
SP 58 & $\ge$4$\text{j}, ~ \ge$2$\cPqb, ~ \pt^{0}, ~ \mTii^{0} ~ $                 & 0.42 $\pm$ 0.06  & 0.18 $\pm$ 0.03  & 0.16 $\pm$ 0.03  & 1.4 $\pm$ 0.1  & 0.18 $\pm$ 0.02  \\
SP 59 & $\ge$4$\text{j}, ~ \ge$2$\cPqb, ~ \pt^{0}, ~ \mTii^{30} ~ $                 & 0.65 $\pm$ 0.07  & 0.26 $\pm$ 0.03  & 0.13 $\pm$ 0.02  & 0.86 $\pm$ 0.08  & 0.35 $\pm$ 0.03  \\
SP 60 & $\ge$4$\text{j}, ~ \ge$2$\cPqb, ~ \pt^{75}, ~ \mTii^{0} ~ $                 & 1.6 $\pm$ 0.3  & 0.67 $\pm$ 0.11  & 0.24 $\pm$ 0.07  & 0.71 $\pm$ 0.08  & 2.4 $\pm$ 0.3  \\
SP 61 & $\ge$4$\text{j}, ~ \ge$2$\cPqb, ~ \pt^{75}, ~ \mTii^{30} ~ $                 & 0.08 $\pm$ 0.02  & 0.03 $\pm$ 0.00  & 0.03 $\pm$ 0.01  & 0.73 $\pm$ 0.07  & 0.44 $\pm$ 0.04  \\
SP 62 & $\ge$4$\text{j}, ~ \ge$2$\cPqb, ~ \pt^{125}, ~ \mTii^{0} ~ $                & 0.14 $\pm$ 0.03  & 0.05 $\pm$ 0.02  & 0.03 $\pm$ 0.00  & 0.53 $\pm$ 0.06  & 0.82 $\pm$ 0.09  \\
SP 63 & $\ge$4$\text{j}, ~ \ge$2$\cPqb, ~ \pt^{125}, ~ \mTii^{30} ~ $                & 0.51 $\pm$ 0.11  & 0.20 $\pm$ 0.06  & 0.11 $\pm$ 0.03  & 0.57 $\pm$ 0.05  & 6.4 $\pm$ 0.6  \\
\hline
\end{tabular}
}
\end{table*}

\begin{table*}[htb]
\centering
\topcaption{The expected signal yields for the SUSY simplified model signals considered
are shown for each search region bin of the EWP analysis. The category that each search
region bin belongs to is also indicated in the table.
The search region bins definitions are summarized in Table~\ref{tab:BinsRazor}.
The labels for the different signal models are explained in detail in the caption of Table~\ref{tab:signalYieldTableB_1}.
}
\cmsTable{
\begin{tabular}{ccccccc}
\hline
Search     & \multirow{2}{*}{Category} & \multirow{2}{*}{$\PH\PH$} & \multirow{2}{*}{$\PZ\PH$}  & \multirow{2}{*}{$\PW\PH$ (200,1)} & \multirow{2}{*}{\PSQb (450,1)} & \multirow{2}{*}{\PSQb (450,300)}  \\
region bin &                           & & & & &    \\
\hline
EWP 0 & Two-Lepton                        &      0.2 $\pm$ 0.01    &   1.6 $\pm$   0.1    &    0.0 $\pm$ 0.000     &       0.2 $\pm$   0.1    &        0.1 $\pm$ 0.03       \\
EWP 1 & Muon High-$\pt$                      &      4.5 $\pm$  0.2    &   1.5 $\pm$   0.1    &    3.3 $\pm$   0.2     &       4.4 $\pm$   1.8    &        0.9 $\pm$  0.4       \\
EWP 2 & Muon Low-$\pt$                       &      1.6 $\pm$ 0.04    &   0.6 $\pm$  0.02    &    1.7 $\pm$  0.05     &       0.6 $\pm$   0.2    &        1.8 $\pm$  0.7       \\
EWP 3 & Electron High-$\pt$                  &      4.0 $\pm$  0.2    &   1.5 $\pm$   0.1    &    2.7 $\pm$   0.1     &       3.2 $\pm$   1.3    &        0.8 $\pm$  0.3       \\
EWP 4 & Electron Low-$\pt$                   &      0.5 $\pm$ 0.01    &   0.2 $\pm$  0.01    &    0.9 $\pm$  0.04     &       0.1 $\pm$  0.03    &        0.7 $\pm$  0.3       \\
EWP 5 & Electron Low-$\pt$                   &      0.3 $\pm$ 0.01    &   0.1 $\pm$  0.01    &    0.2 $\pm$  0.02     &       0.2 $\pm$   0.1    &        0.2 $\pm$  0.1       \\
EWP 6 & Electron Low-$\pt$                   &      0.3 $\pm$ 0.01    &   0.2 $\pm$ 0.004    &    0.3 $\pm$  0.02     &       0.1 $\pm$  0.04    &        0.4 $\pm$  0.2       \\
EWP 7 & $\PH\PQb\cPaqb$ High-$\pt$  &     11.9 $\pm$  0.5    &   3.4 $\pm$   0.2    &    0.2 $\pm$  0.01     &       4.3 $\pm$   4.3    &        4.7 $\pm$  1.9       \\
EWP 8 & $\PH\PQb\cPaqb$ High-$\pt$  &      9.1 $\pm$  0.6    &   2.5 $\pm$   0.2    &    0.1 $\pm$ 0.005     &      30.1 $\pm$  12.1    &        2.2 $\pm$  0.8       \\
EWP 9 & $\PH\PQb\cPaqb$ Low-$\pt$   &      1.9 $\pm$  0.2    &   0.6 $\pm$  0.05    &    0.1 $\pm$ 0.003     &       0.8 $\pm$   1.0    &        6.5 $\pm$  2.8       \\
EWP 10 & $\PH\PQb\cPaqb$ Low-$\pt$  &      1.2 $\pm$  0.1    &   0.4 $\pm$  0.04    &   0.03 $\pm$ 0.002     &       3.7 $\pm$   1.5    &        2.4 $\pm$  1.0       \\
EWP 11 & $\PZ\PQb\cPaqb$ High-$\pt$ &      3.2 $\pm$  0.3    &   1.7 $\pm$   0.2    &    0.3 $\pm$  0.02     &       0.6 $\pm$   0.6    &        1.9 $\pm$  0.8       \\
EWP 12 & $\PZ\PQb\cPaqb$ High-$\pt$ &      1.3 $\pm$  0.2    &   0.6 $\pm$   0.1    &    0.1 $\pm$  0.01     &       4.8 $\pm$   2.2    &        0.4 $\pm$  0.2       \\
EWP 13 & $\PZ\PQb\cPaqb$ High-$\pt$ &      2.5 $\pm$  0.1    &   1.1 $\pm$   0.1    &    0.1 $\pm$  0.02     &       2.3 $\pm$   2.2    &        1.0 $\pm$  0.4       \\
EWP 14 & $\PZ\PQb\cPaqb$ Low-$\pt$  &      1.7 $\pm$  0.2    &   0.8 $\pm$   0.1    &    0.2 $\pm$  0.01     &       0.1 $\pm$   0.1    &        3.7 $\pm$  1.5       \\
EWP 15 & $\PZ\PQb\cPaqb$ Low-$\pt$  &      0.6 $\pm$  0.2    &   0.2 $\pm$  0.04    &   0.02 $\pm$ 0.002     &       0.6 $\pm$   0.3    &        0.8 $\pm$  0.4       \\
EWP 16 & $\PZ\PQb\cPaqb$ Low-$\pt$  &      1.0 $\pm$ 0.05    &   0.4 $\pm$  0.02    &   0.04 $\pm$  0.01     &       0.3 $\pm$   0.3    &        1.5 $\pm$  0.6       \\
EWP 17 & High-$\pt$                          &      5.3 $\pm$  1.6    &   5.5 $\pm$   0.6    &    7.2 $\pm$   0.5     &       0.3 $\pm$   0.2    &        1.4 $\pm$  0.7       \\
EWP 18 & High-$\pt$                          &      1.8 $\pm$  0.1    &   0.8 $\pm$  0.05    &    0.5 $\pm$  0.03     &      0.01 $\pm$   0.1    &        0.3 $\pm$  0.1       \\
EWP 19 & High-$\pt$                          &      6.0 $\pm$  1.4    &   4.0 $\pm$   0.7    &    3.6 $\pm$   0.2     &       0.6 $\pm$   0.4    &        1.4 $\pm$  0.6       \\
EWP 20 & High-$\pt$                          &     42.1 $\pm$  3.9    &  19.6 $\pm$   1.8    &    9.1 $\pm$   0.8     &      40.1 $\pm$  15.8    &        6.1 $\pm$  2.4       \\
EWP 21 & High-$\pt$                          &      4.9 $\pm$  0.2    &   2.3 $\pm$   0.1    &    1.4 $\pm$   0.1     &      0.03 $\pm$  0.04    &        0.9 $\pm$  0.4       \\
EWP 22 & High-$\pt$                          &      7.3 $\pm$  1.2    &   4.2 $\pm$   0.6    &    3.0 $\pm$   0.2     &       1.5 $\pm$   1.4    &        1.3 $\pm$  0.5       \\
EWP 23 & High-Res                         &      1.1 $\pm$  1.2    &   1.0 $\pm$   0.4    &    3.0 $\pm$   0.6     &      0.03 $\pm$  0.02    &        2.2 $\pm$  1.2       \\
EWP 24 & High-Res                         &      1.5 $\pm$  0.5    &   0.9 $\pm$   0.2    &    1.1 $\pm$   0.1     &      0.03 $\pm$  0.01    &        1.4 $\pm$  0.6       \\
EWP 25 & High-Res                         &      0.6 $\pm$  0.3    &   0.4 $\pm$   0.1    &    0.6 $\pm$   0.1     &      0.01 $\pm$   0.2    &        0.6 $\pm$  0.3       \\
EWP 26 & High-Res                         &     13.7 $\pm$  2.1    &   6.5 $\pm$   1.0    &    4.4 $\pm$   0.7     &       4.1 $\pm$   1.7    &       10.4 $\pm$  4.4       \\
EWP 27 & High-Res                         &      0.5 $\pm$  0.1    &   0.3 $\pm$  0.04    &    0.2 $\pm$  0.03     &       0.0 $\pm$ 0.000    &        0.4 $\pm$  0.2       \\
EWP 28 & High-Res                         &      0.8 $\pm$  0.2    &   0.5 $\pm$   0.1    &    0.6 $\pm$   0.1     &       0.1 $\pm$   0.2    &        0.9 $\pm$  0.4       \\
EWP 29 & Low-Res                          &      0.7 $\pm$  0.7    &   0.7 $\pm$   0.3    &    1.9 $\pm$   0.5     &      0.02 $\pm$  0.01    &        1.5 $\pm$  0.8       \\
EWP 30 & Low-Res                          &      1.0 $\pm$  0.3    &   0.5 $\pm$   0.1    &    0.7 $\pm$   0.2     &      0.02 $\pm$  0.01    &        1.0 $\pm$  0.5       \\
EWP 31 & Low-Res                          &      0.5 $\pm$  0.4    &   0.3 $\pm$   0.2    &    0.4 $\pm$   0.1     &      0.01 $\pm$ 0.003    &        0.5 $\pm$  0.3       \\
EWP 32 & Low-Res                          &      8.4 $\pm$  2.2    &   4.1 $\pm$   1.0    &    3.0 $\pm$   0.8     &       2.7 $\pm$   1.3    &        7.1 $\pm$  3.6       \\
EWP 33 & Low-Res                          &      0.4 $\pm$  0.1    &   0.2 $\pm$  0.05    &    0.2 $\pm$   0.04    &     0.002 $\pm$ 0.001    &        0.2 $\pm$  0.1       \\
EWP 34 & Low-Res                          &      0.6 $\pm$  0.2    &   0.3 $\pm$   0.1    &    0.4 $\pm$   0.1     &      0.01 $\pm$  0.01    &        0.6 $\pm$  0.3       \\

\hline
\end{tabular}
}
\label{tab:signalYieldTableA}
\end{table*}

\begin{figure*}[hbtp]
\centering
  \includegraphics[width=0.6\textwidth,angle=0.]{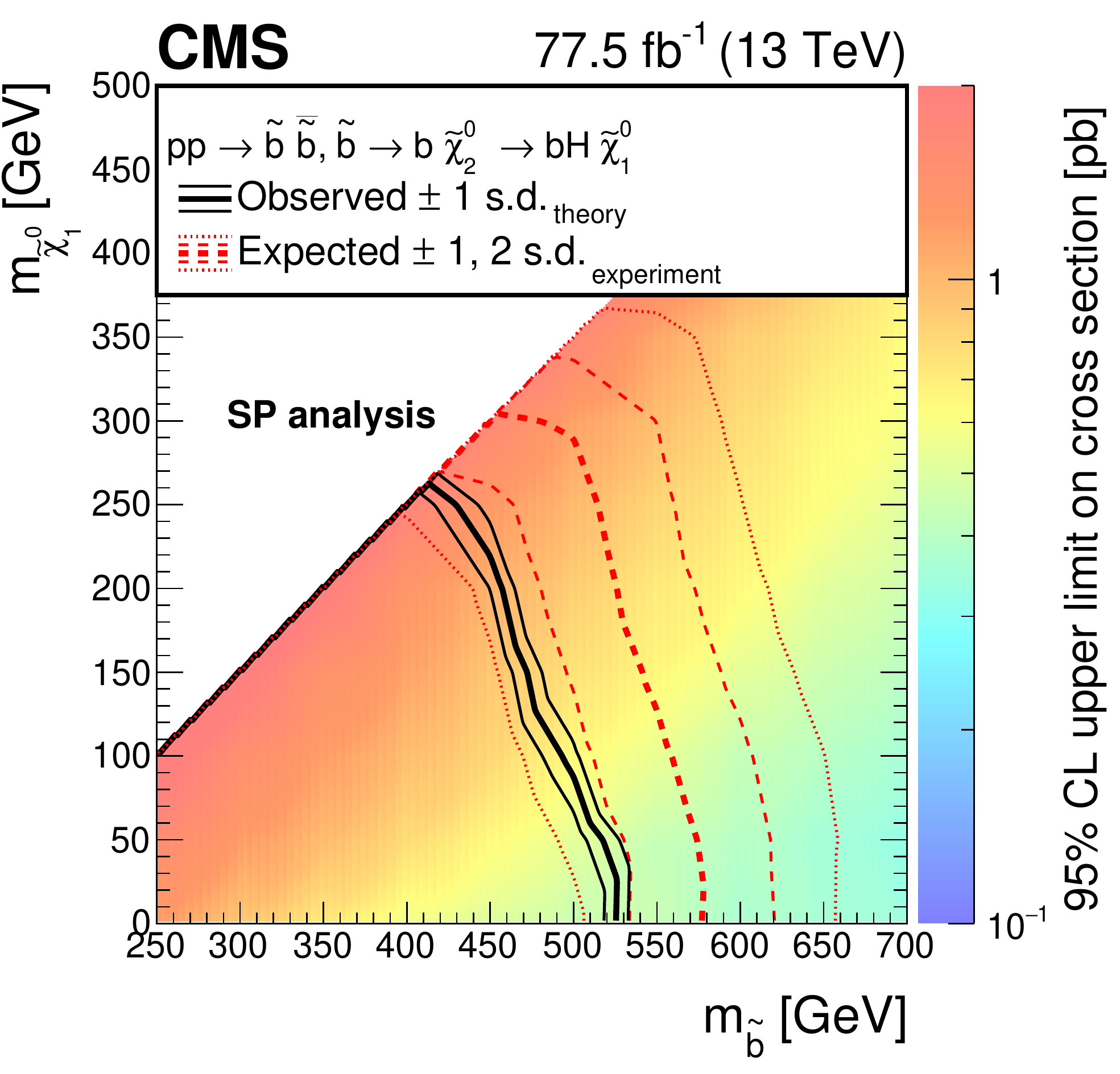}
  \caption{The observed 95\% \CL upper limits on the bottom squark pair production cross section are shown for the SP analysis. The bold and light solid black contours represent the observed exclusion region and the $\pm$1 standard deviation (s.d.) band, including both experimental and theoretical uncertainties. The analogous red dotted contours represent the expected exclusion region and its $\pm$1 and $\pm$2 s.d. bands.
}
    \label{fig:LimitsSbottom}
\end{figure*}

Following the \CLs criterion~\cite{Junk:1999kv,Read:2002hq,ATLAS:2011tau}, we use
the profile likelihood ratio test statistic and the asymptotic
formula~\cite{Cowan:2010js} to evaluate the 95\%~confidence level (\CL) observed and expected limits on
the signal production cross sections. For the simplified models of bottom squark pair production
where the bottom squark undergoes a cascade decay to a Higgs boson and the LSP,
the SP analysis yields better expected sensitivity because of the binning in the number of jets and \cPqb-tagged jets,
as more jets and more heavy-flavor jets are produced. The limits obtained using the SP analysis
are shown in Fig.~\ref{fig:LimitsSbottom}, as a function of the
bottom squark mass and the LSP mass. We exclude bottom squarks with masses
below about 530\GeV for an LSP mass of 1\GeV.

\begin{figure*}[hbtp]
\centering
  \includegraphics[width=0.6\textwidth,angle=0.]{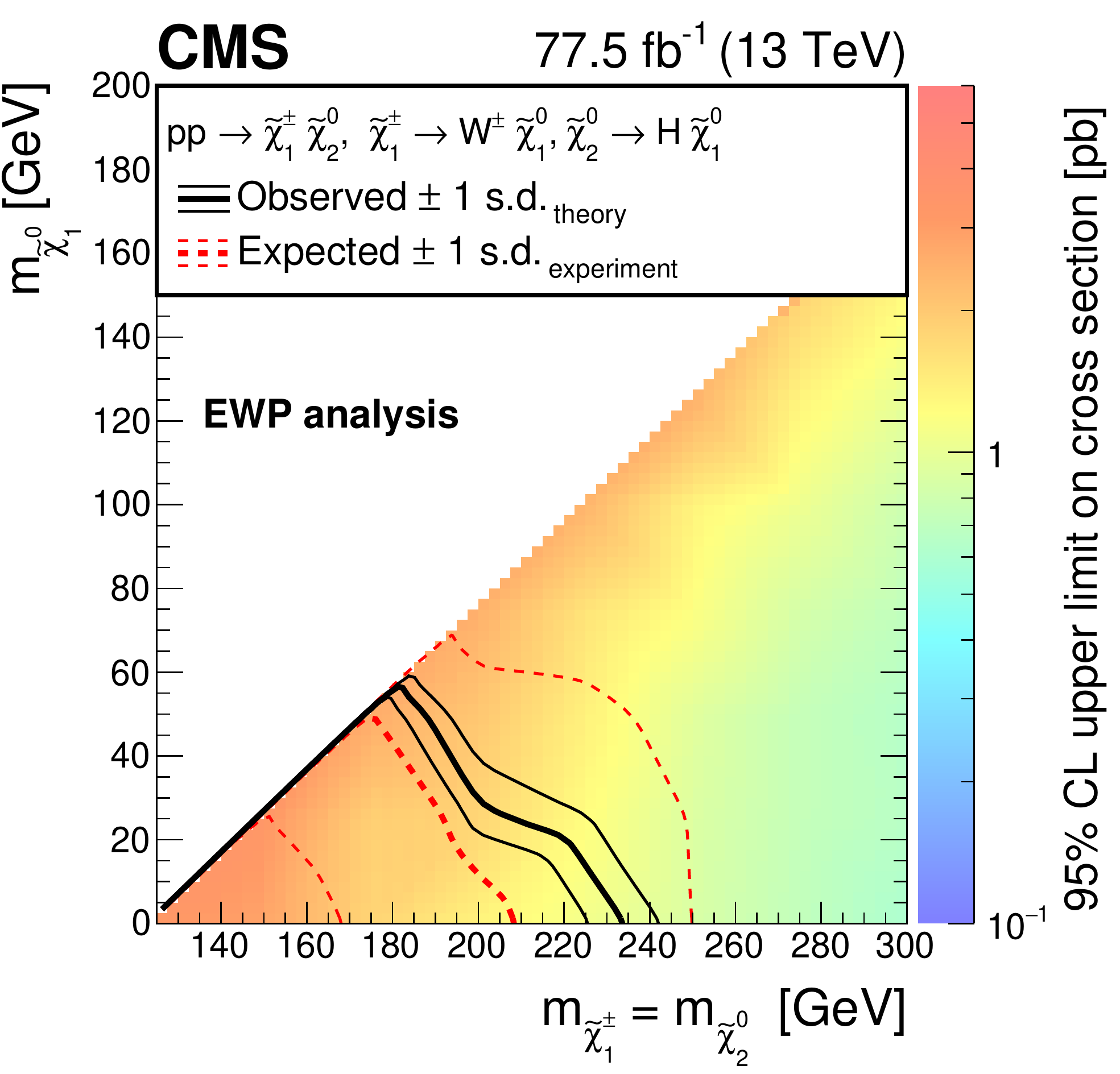}

  \caption{The observed 95\% \CL upper limits on the wino-like chargino-neutralino production cross section are shown for the EWP analysis. The bold and light black contours represent the observed exclusion region and the $\pm$1 standard deviation (s.d.) band, including both experimental and theoretical uncertainties. The analogous red dotted contours represent the expected exclusion region and its $\pm$1 s.d. band.
 }
    \label{fig:LimitsTChiWH}
\end{figure*}

\begin{figure*}[hbtp]
\centering
  \includegraphics[width=0.49\textwidth,angle=0.]{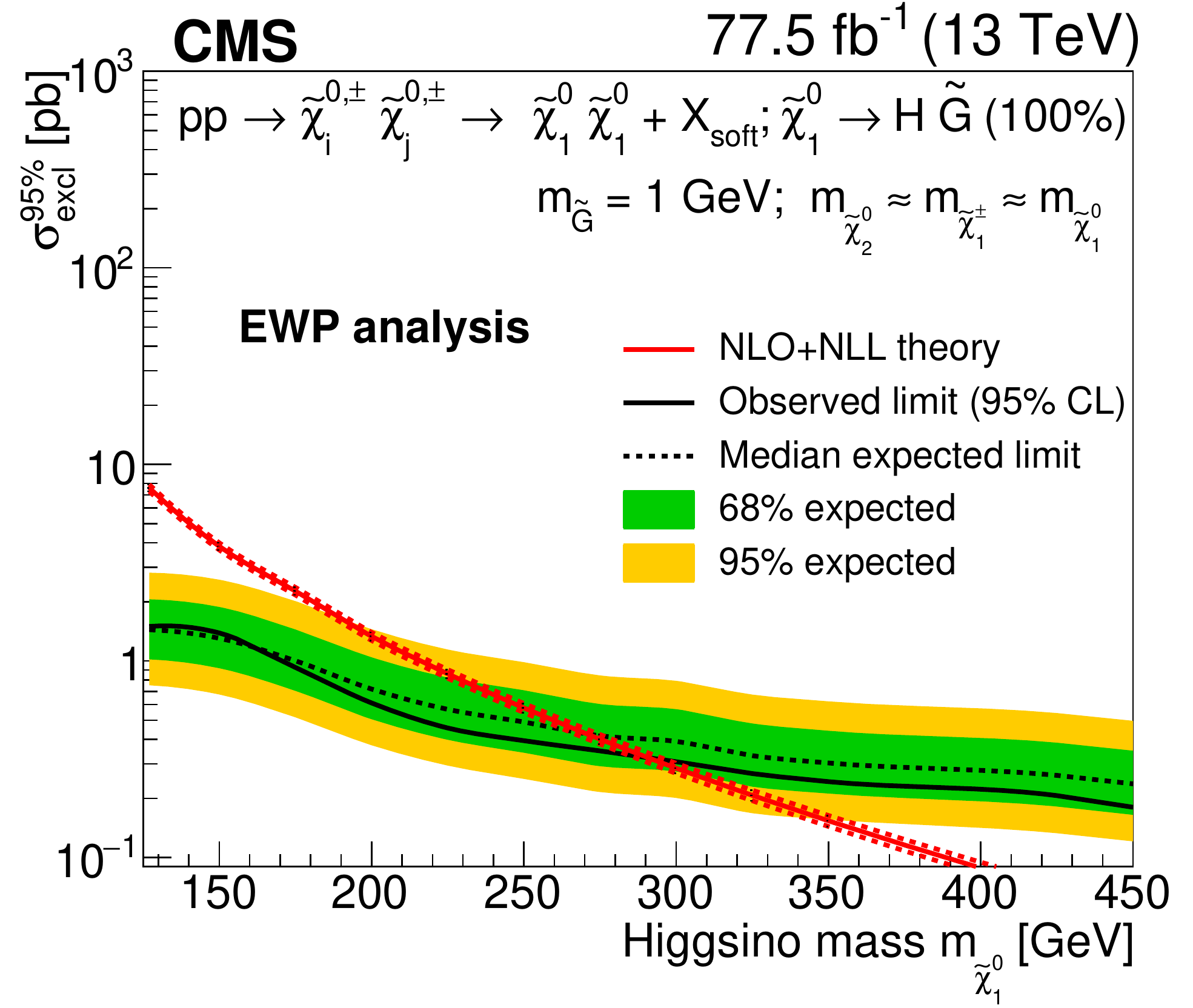}
  \includegraphics[width=0.49\textwidth,angle=0.]{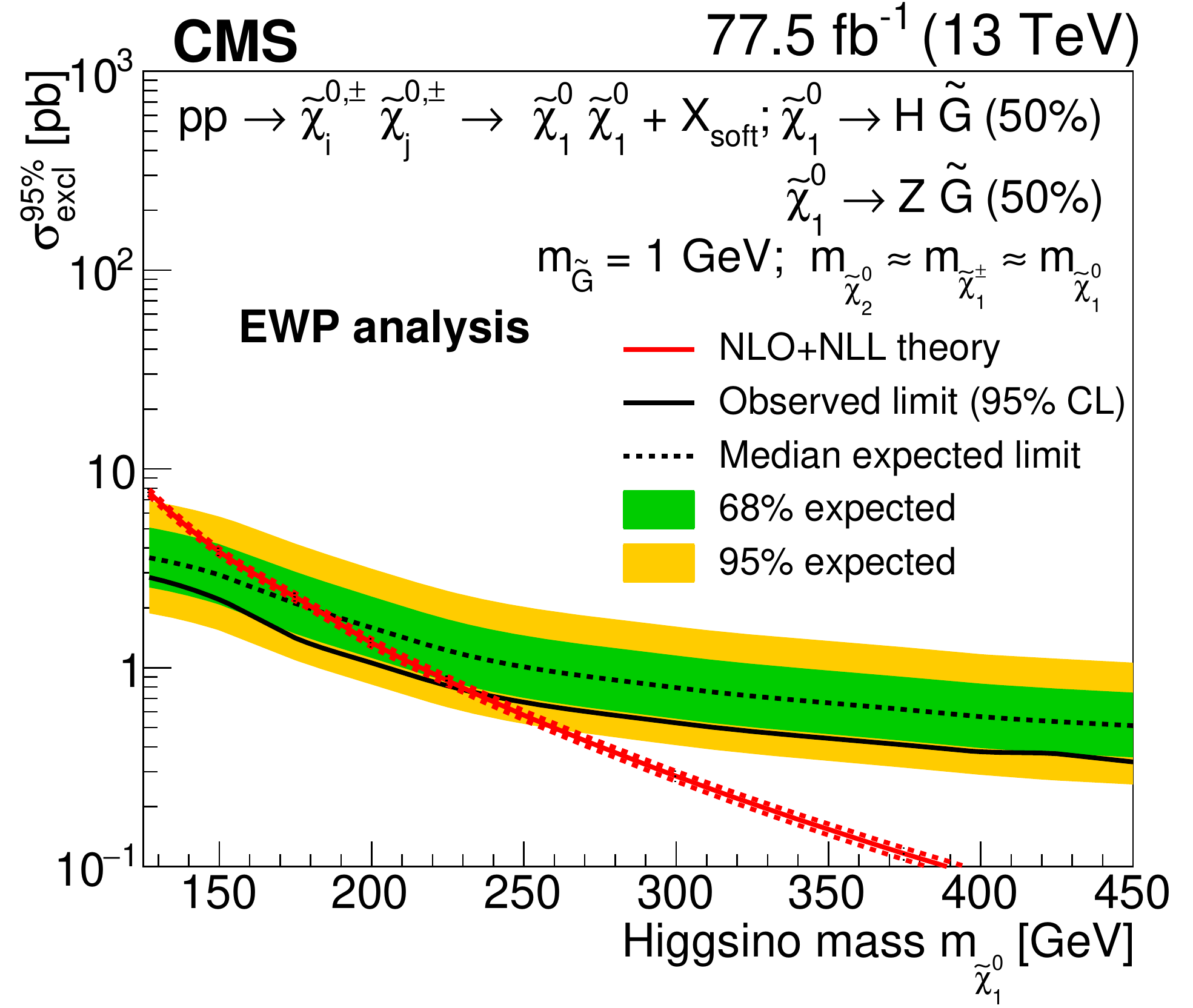}
\caption{The observed 95\% \CL upper limits on the production cross section for higgsino-like chargino-neutralino
        production are shown for the EWP analysis. We present limits in the scenario where the branching fraction of $\PSGczDo\to \PH\PXXSG$ decay is 100\% (left plot), and where the $\PSGczDo\to \PH\PXXSG$ and $\PSGczDo\to \PZ\PXXSG$ decays are each 50\% (right plot). The dotted and solid black curves represent the expected and observed exclusion region, and the green dark and yellow light bands represent the $\pm$1 and $\pm$2 standard deviation regions, respectively. The red solid and dotted lines show the theoretical production cross section and its uncertainty band. \label{fig:LimitsTChiHHnHZ}}
\end{figure*}

For the simplified models of chargino-neutralino production, the EWP analysis has slightly better
expected sensitivity because of the inclusion of bins with smaller \MR \,and larger \Rtwo.
Events in such bins typically have lower values of \ptmiss and are not in the regions of
high signal sensitivity for the SP analysis, while the \Rtwo variable is able to suppress
backgrounds more effectively in these regions of phase space.
For the wino-like chargino-neutralino production, the limits obtained using the EWP analysis are
shown in Fig.~\ref{fig:LimitsTChiWH} as a function of the chargino mass and the LSP mass.
We exclude chargino masses below about 235\GeV for an LSP mass of 1\GeV. For the higgsino-like
chargino-neutralino production simplified models, the limits obtained using the EWP analysis
are shown in Fig.~\ref{fig:LimitsTChiHHnHZ} as a function of the chargino mass for the case
where the branching fraction of the $\PSGczDo\to \PH\PXXSG$ decay is 100\%, and for
the case where the branching fraction of the $\PSGczDo\to \PH\PXXSG$ and
$\PSGczDo\to \PZ\PXXSG$ decays are both 50\%.
We exclude charginos below 290 and 230 \GeV in the former and latter cases, respectively.
The corresponding limits from the EWP analysis as applied to bottom
squark production and limits from the SP analysis as applied to
chargino-neutralino production are included in the appendix for
completeness.

The search region bins with large $\pt^{\gamma\gamma}$ in the \hbb category yield the best
overall sensitivity. For signal models with squark or neutralino masses exceeding the
Higgs boson mass by $100$~GeV or more, the search region bins with large values
of $\pt^{\gamma\gamma}$ and large values of the kinematic variables \MR~and \mTii 
in the untagged jet categories of the SP analysis or the High-$\pt$ category for the EWP 
analysis also contribute significantly to the search sensitivity. 
For more compressed regions of the signal model parameter space, where the 
neutralino mass approaches the Higgs boson mass, the search region bins with 
large $\pt^{\gamma\gamma}$ in the leptonic categories contribute significantly
to the search sensitivity. The search region bins with small values 
of $\pt^{\gamma\gamma}$ and small values of the kinematic variables \MR, \Rtwo, and \mTii 
typically have low sensitivity to the simplified models considered due to 
higher levels of background, but are included to maintain inclusivity for this search.

\section{Summary}
\label{sec:summary}

We have presented a search for supersymmetry (SUSY) in the final state with a Higgs boson ($\PH$)
decaying to a photon pair, using data collected with the
CMS detector at the LHC in 2016 and 2017, corresponding to 77.5\fbinv of
integrated luminosity. To improve the sensitivity over previously published
results, we pursue two strategies that are optimized
for strong and electroweak SUSY production, respectively.
Photon pairs in the central region of the detector are used
to reconstruct Higgs boson candidates. Charged leptons and \cPqb jets are used to tag the decay products of an additional boson, while
kinematic quantities such as $\mTii$ and the razor variables $\MR$ and $\Rtwo$ are used to suppress standard model backgrounds. Data
driven fits determine the shape and normalization of the nonresonant background. The resonant background from standard model Higgs boson production is
estimated from simulation. The results are interpreted in terms of exclusion
limits on the production cross section of simplified models of bottom squark pair production
and chargino-neutralino production. As a result of the improvements in the event categorization
and the larger data set, we extend the mass limits over the previous best CMS 
results~\cite{Sirunyan:2017eie,Aaboud:2018ngk} by about 100\GeV for bottom squark pair production and
about 50\GeV for chargino-neutralino production. We exclude bottom squark pair production for
bottom squark masses below 530\GeV for a lightest neutralino mass of 1\GeV;
wino-like chargino-neutralino production, for chargino and neutralino ($\PSGczDo$) masses of up to 235\GeV and a gravitino (\PXXSG) mass of 1\GeV;
and higgsino-like chargino-neutralino production, for chargino and neutralino ($\PSGczDo$) masses of up to 290 and 230\GeV for the cases where the
branching fraction of the lightest neutralino $\PSGczDo\to \PH\PXXSG$ decay is 100\%, and where the branching fractions of the $\PSGczDo\to\PH\PXXSG$
and $\PSGczDo\to\PZ\PXXSG$ decays are both 50\%, respectively.

\clearpage

\begin{acknowledgments}
We congratulate our colleagues in the CERN accelerator departments for the excellent performance of the LHC and thank the technical and administrative staffs at CERN and at other CMS institutes for their contributions to the success of the CMS effort. In addition, we gratefully acknowledge the computing centers and personnel of the Worldwide LHC Computing Grid for delivering so effectively the computing infrastructure essential to our analyses. Finally, we acknowledge the enduring support for the construction and operation of the LHC and the CMS detector provided by the following funding agencies: BMBWF and FWF (Austria); FNRS and FWO (Belgium); CNPq, CAPES, FAPERJ, FAPERGS, and FAPESP (Brazil); MES (Bulgaria); CERN; CAS, MoST, and NSFC (China); COLCIENCIAS (Colombia); MSES and CSF (Croatia); RPF (Cyprus); SENESCYT (Ecuador); MoER, ERC IUT, PUT and ERDF (Estonia); Academy of Finland, MEC, and HIP (Finland); CEA and CNRS/IN2P3 (France); BMBF, DFG, and HGF (Germany); GSRT (Greece); NKFIA (Hungary); DAE and DST (India); IPM (Iran); SFI (Ireland); INFN (Italy); MSIP and NRF (Republic of Korea); MES (Latvia); LAS (Lithuania); MOE and UM (Malaysia); BUAP, CINVESTAV, CONACYT, LNS, SEP, and UASLP-FAI (Mexico); MOS (Montenegro); MBIE (New Zealand); PAEC (Pakistan); MSHE and NSC (Poland); FCT (Portugal); JINR (Dubna); MON, RosAtom, RAS, RFBR, and NRC KI (Russia); MESTD (Serbia); SEIDI, CPAN, PCTI, and FEDER (Spain); MOSTR (Sri Lanka); Swiss Funding Agencies (Switzerland); MST (Taipei); ThEPCenter, IPST, STAR, and NSTDA (Thailand); TUBITAK and TAEK (Turkey); NASU and SFFR (Ukraine); STFC (United Kingdom); DOE and NSF (USA).

\hyphenation{Rachada-pisek} Individuals have received support from the Marie-Curie program and the European Research Council and Horizon 2020 Grant, contract Nos.\ 675440, 752730, and 765710 (European Union); the Leventis Foundation; the A.P.\ Sloan Foundation; the Alexander von Humboldt Foundation; the Belgian Federal Science Policy Office; the Fonds pour la Formation \`a la Recherche dans l'Industrie et dans l'Agriculture (FRIA-Belgium); the Agentschap voor Innovatie door Wetenschap en Technologie (IWT-Belgium); the F.R.S.-FNRS and FWO (Belgium) under the ``Excellence of Science -- EOS" -- be.h project n.\ 30820817; the Beijing Municipal Science \& Technology Commission, No. Z181100004218003; the Ministry of Education, Youth and Sports (MEYS) of the Czech Republic; the Lend\"ulet (``Momentum") Program and the J\'anos Bolyai Research Scholarship of the Hungarian Academy of Sciences, the New National Excellence Program \'UNKP, the NKFIA research grants 123842, 123959, 124845, 124850, 125105, 128713, 128786, and 129058 (Hungary); the Council of Science and Industrial Research, India; the HOMING PLUS program of the Foundation for Polish Science, cofinanced from European Union, Regional Development Fund, the Mobility Plus program of the Ministry of Science and Higher Education, the National Science Center (Poland), contracts Harmonia 2014/14/M/ST2/00428, Opus 2014/13/B/ST2/02543, 2014/15/B/ST2/03998, and 2015/19/B/ST2/02861, Sonata-bis 2012/07/E/ST2/01406; the National Priorities Research Program by Qatar National Research Fund; the Ministry of Science and Education, grant no. 3.2989.2017 (Russia); the Programa Estatal de Fomento de la Investigaci{\'o}n Cient{\'i}fica y T{\'e}cnica de Excelencia Mar\'{\i}a de Maeztu, grant MDM-2015-0509 and the Programa Severo Ochoa del Principado de Asturias; the Thalis and Aristeia programs cofinanced by EU-ESF and the Greek NSRF; the Rachadapisek Sompot Fund for Postdoctoral Fellowship, Chulalongkorn University and the Chulalongkorn Academic into Its 2nd Century Project Advancement Project (Thailand); the Welch Foundation, contract C-1845; and the Weston Havens Foundation (USA).
\end{acknowledgments}

\bibliography{auto_generated}

\clearpage

\appendix

\section{Additional simplified model interpretations}

While the EWP and SP analyses have greater expected sensitivity to electroweak and strong SUSY production, respectively,
both analyses do have sensitivity to both production modes. In this appendix, we present
limits obtained from the EWP and SP analyses for the simplified models that were not shown in
Section~\ref{sec:results}.

\begin{figure*}[hbtp]
\centering
  \includegraphics[width=0.6\textwidth,angle=0.]{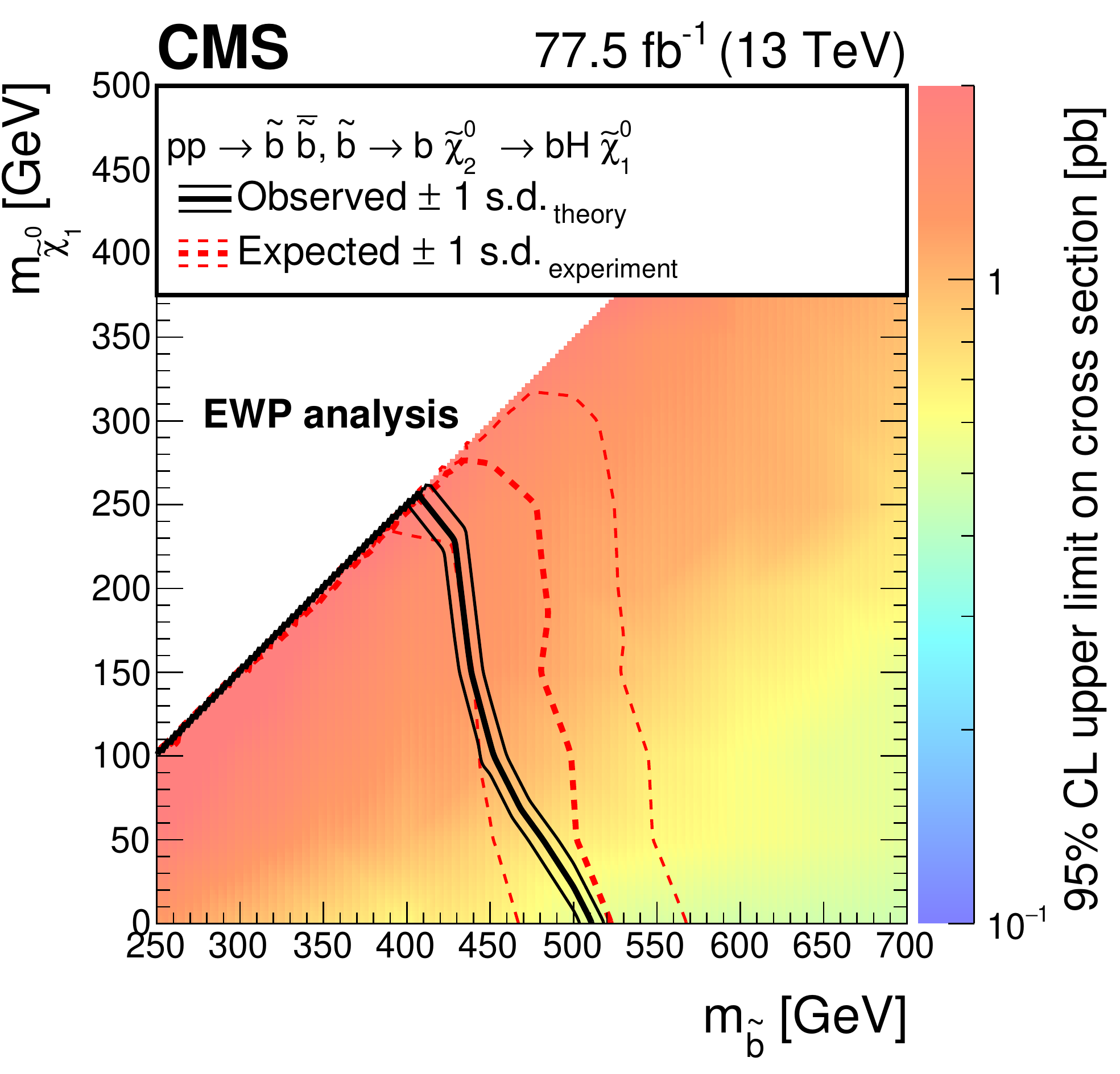}
  \includegraphics[width=0.6\textwidth,angle=0.]{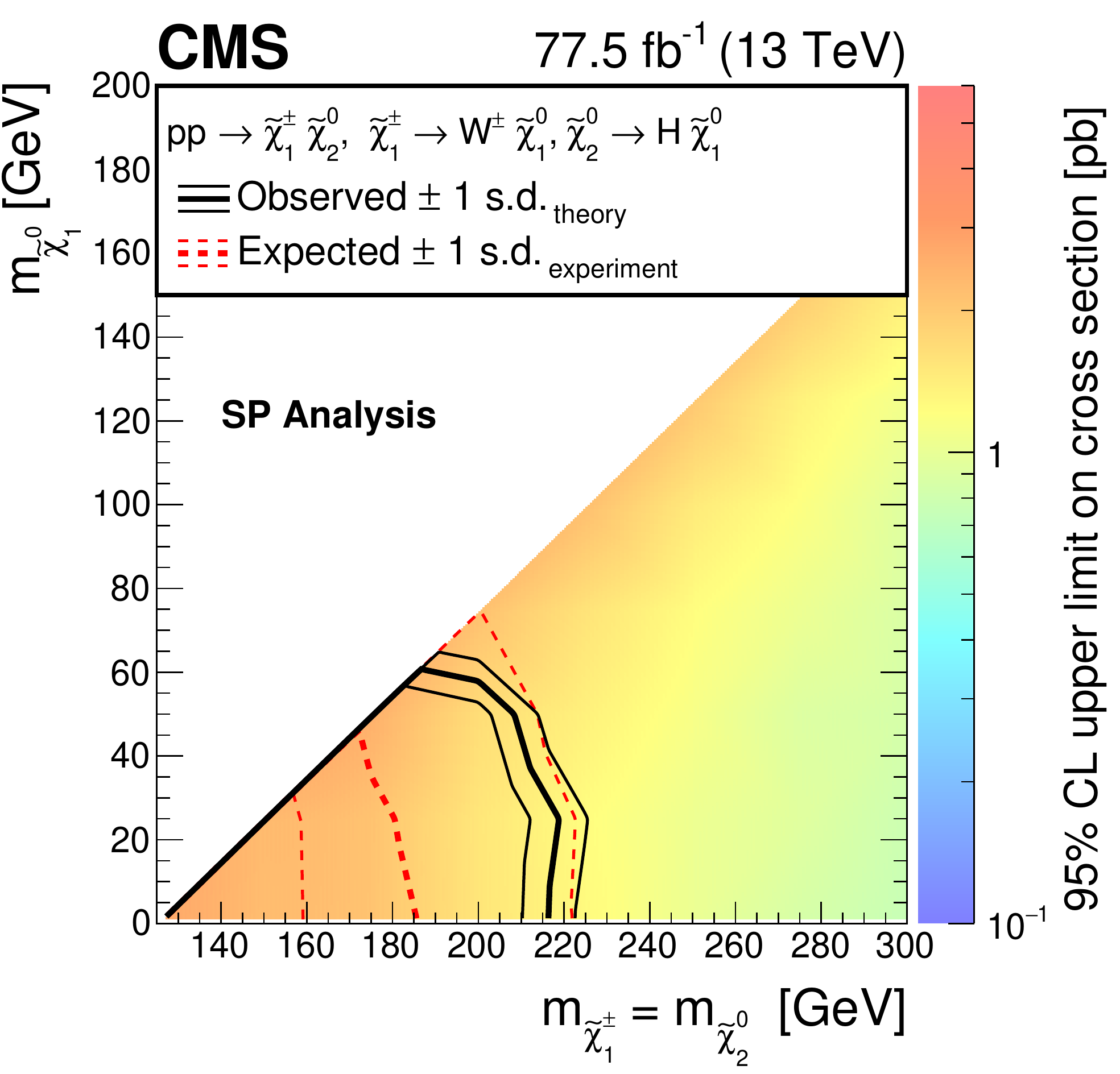}
  \caption{The observed 95\% \CL upper limits on the bottom squark pair production cross section for
the EWP analysis (upper plot), and on the wino-like chargino-neutralino production cross
section for the SP analysis (lower plot), are shown. The bold and light solid black contours
represent the observed exclusion region and the $\pm$1 standard deviation (s.d.) band, including
both experimental and theoretical uncertainties. The analogous red dotted contours represent the
expected exclusion region and its $\pm$1 s.d. band.
}
    \label{fig:LimitsSbottomEWP_TChiWH_SP}
\end{figure*}

The upper plot of Figure~\ref{fig:LimitsSbottomEWP_TChiWH_SP} shows the limits for sbottom pair production
obtained using the EWP analysis, as a function of the
bottom squark mass and the LSP mass.

\begin{figure*}[hbtp]
\centering
  \includegraphics[width=0.49\textwidth,angle=0.]{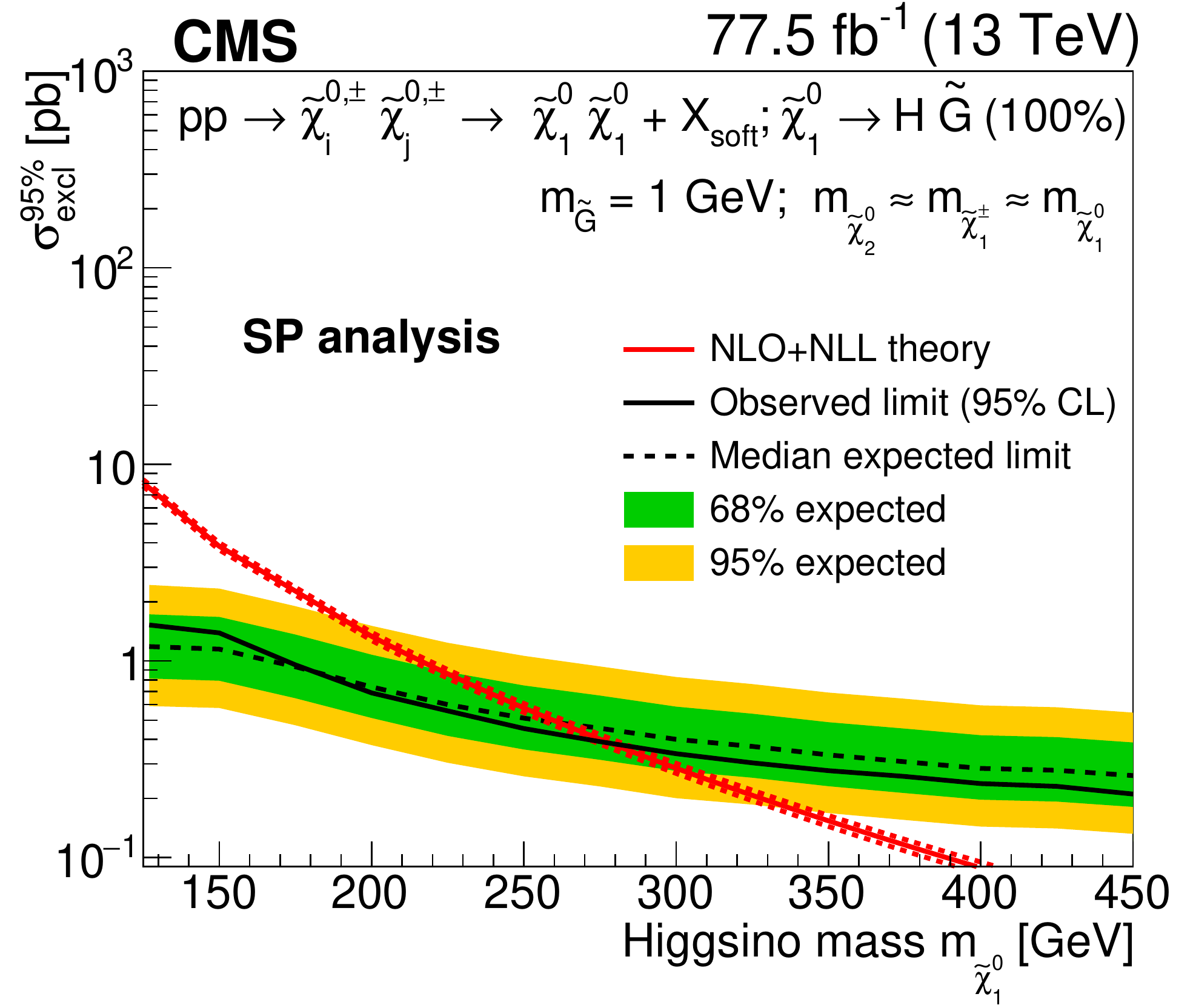}
  \includegraphics[width=0.49\textwidth,angle=0.]{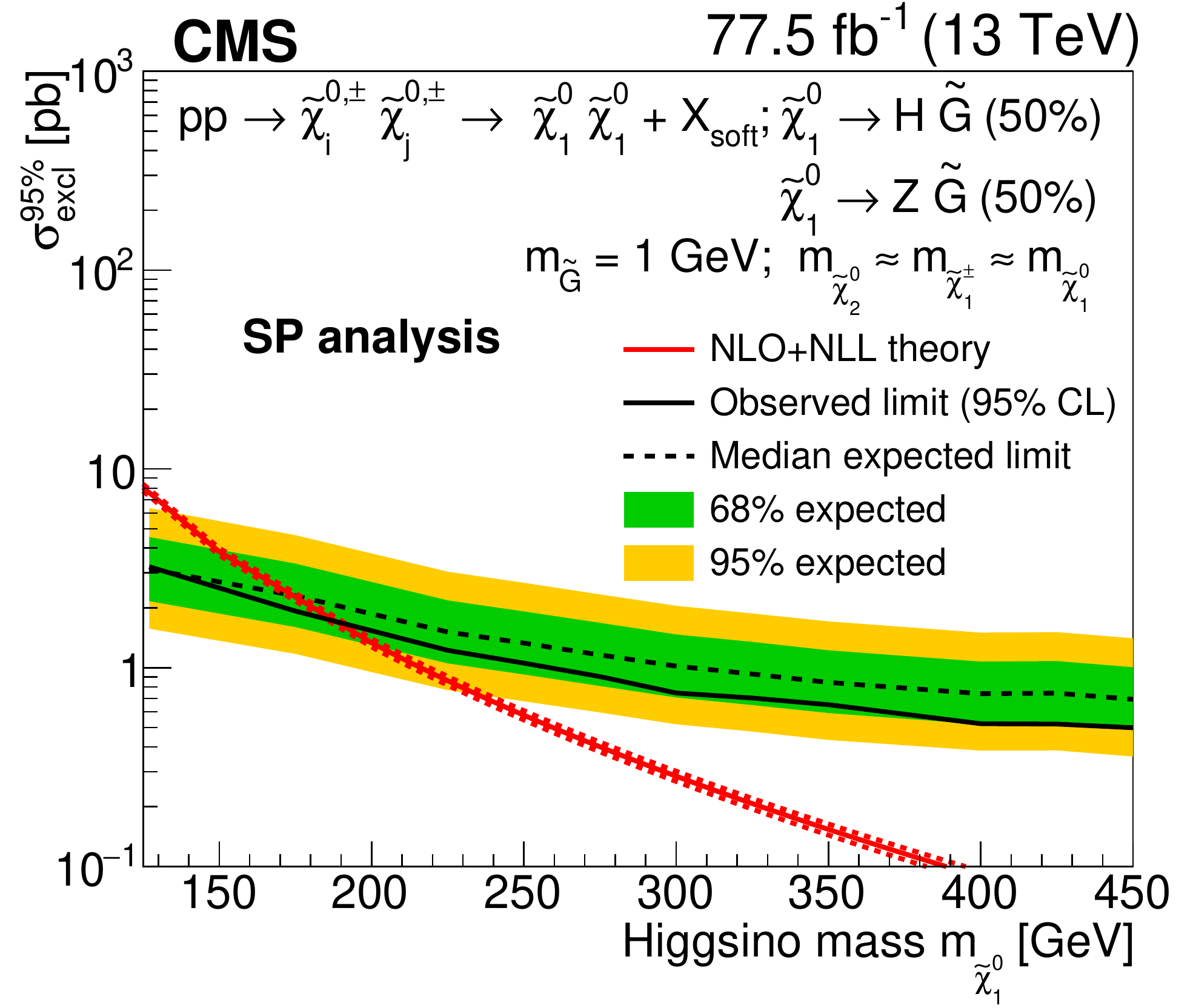}
\caption{The observed 95\% \CL upper limits on the production cross section for higgsino-like chargino-neutralino
        production are shown for the SP analysis. The charginos and neutralinos undergo several cascade decays producing either Higgs bosons (left plot), or a Higgs boson and a $\PZ$ boson (right plot). We present limits in the scenario where the branching fraction of $\PSGczDo\to \PH\PXXSG$ decay is 100\% (left plot), and where the $\PSGczDo\to \PH\PXXSG$ and $\PSGczDo\to \PZ\PXXSG$ decays are each 50\% (right plot). The dotted and solid black curves represent the expected and observed exclusion region, and the green dark and yellow light bands represent the $\pm$1 and $\pm$2 standard deviation regions, respectively. The red solid and dotted lines show the theoretical production cross section and its uncertainty band. \label{fig:LimitsTChiHHnHZ_SP}}
\end{figure*}

For the wino-like chargino-neutralino production, the limits obtained using the SP analysis are
shown in the lower plot of Fig.~\ref{fig:LimitsSbottomEWP_TChiWH_SP} as a function of the chargino mass and the LSP mass.
Figure~\ref{fig:LimitsTChiHHnHZ_SP} shows the limits for
the higgsino-like chargino-neutralino production simplified models obtained using the SP analysis
as a function of the chargino mass for the case
where the branching fraction of the $\PSGczDo\to \PH\PXXSG$ decay is 100\% on the left, and for
the case where the branching fraction of the $\PSGczDo\to \PH\PXXSG$ and
$\PSGczDo\to \PZ\PXXSG$ decays are both 50\% on the right.

\cleardoublepage \section{The CMS Collaboration \label{app:collab}}\begin{sloppypar}\hyphenpenalty=5000\widowpenalty=500\clubpenalty=5000\vskip\cmsinstskip
\textbf{Yerevan Physics Institute, Yerevan, Armenia}\\*[0pt]
A.M.~Sirunyan$^{\textrm{\dag}}$, A.~Tumasyan
\vskip\cmsinstskip
\textbf{Institut f\"{u}r Hochenergiephysik, Wien, Austria}\\*[0pt]
W.~Adam, F.~Ambrogi, T.~Bergauer, J.~Brandstetter, M.~Dragicevic, J.~Er\"{o}, A.~Escalante~Del~Valle, M.~Flechl, R.~Fr\"{u}hwirth\cmsAuthorMark{1}, M.~Jeitler\cmsAuthorMark{1}, N.~Krammer, I.~Kr\"{a}tschmer, D.~Liko, T.~Madlener, I.~Mikulec, N.~Rad, J.~Schieck\cmsAuthorMark{1}, R.~Sch\"{o}fbeck, M.~Spanring, D.~Spitzbart, W.~Waltenberger, C.-E.~Wulz\cmsAuthorMark{1}, M.~Zarucki
% ...
\vskip\cmsinstskip
\dag: Deceased\\
1:  Also at Vienna University of Technology, Vienna, Austria\\
% ...
\end{sloppypar}
\end{document}